\documentclass[preprint2]{aastex}

\usepackage{rotating}
\usepackage{subfigure}

\newcommand{\moo}{\rm $\mu$m}

\shorttitle{Origin of the MFIR emission in radio galaxies II}
\shortauthors{Dicken et al.}

\begin{document}

\title{The origin of the infrared emission in radio galaxies II: \\ analysis of mid- to far-infrared Spitzer observations of the 2Jy sample}

\author{D. Dicken\altaffilmark{1}, C. Tadhunter\altaffilmark{1}, D. Axon\altaffilmark{2}, R.
Morganti\altaffilmark{3,4}, K. J. Inskip\altaffilmark{5}, J. Holt\altaffilmark{6}, R.Gonz\'alez Delgado\altaffilmark{7}, B. Groves\altaffilmark{6} }

    \altaffiltext{1}{Department of Physics and Astronomy, University of
    Sheffield, Hounsfield Road, Sheffield, S3 7RH; d.dicken@sheffield.ac.uk, c.tadhunter@sheffield.ac.uk} 
    \altaffiltext{2}{Department of Physics and Astronomy, Rochester
    Institute of Technology, 84 Lomb Memorial Drive, Rochester NY
    14623; djasps@rit.edu}
    \altaffiltext{3}{ASTRON, P.O. Box 2,
    7990 AA Dwingeloo Netherlands; morganti@astron.nl}
    \altaffiltext{4}{Kapteyn Astronomical Institute, University of Groningen Postbus 800, 9700 AV Groningen, Netherlands}
    \altaffiltext{5}{Max-Planck-Institut f\"{u}r Astronomie, K\"{o}nigstuhl 17, D-69117 Heidelberg, Germany, inskip@mpia-hd.mpg.de}
    \altaffiltext{6}{Leiden Observatory, Leiden University, P.O. Box 9513, 2300 RA Leiden, Netherlands, jholt@strw.leidenuniv.nl, brent@strw.leidenuniv.nl}
    \altaffiltext{7}{Instituto de Astrofisica de Andalucia (CSIC), Apdto.3004, 18080 Granada, Spain; rosa@iaa.es}

\begin{abstract}
We present an analysis of deep mid- to far-infrared (MFIR) Spitzer
photometric observations of the southern 2Jy sample of powerful radio
sources (0.05 $<$ z $<$ 0.7), conducting a statistical investigation
of the links between radio jet, AGN, starburst activity and MFIR
properties. This is part of an ongoing extensive study of powerful
radio galaxies that benefits from both complete optical emission line
information and a uniquely high detection rate in the far-infrared
(far-IR). We find tight correlations between the MFIR and
[OIII]$\lambda$5007 emission luminosities, which are significantly
better than those between MFIR and extended radio luminosities, or
between radio and [OIII] luminosities. Since [OIII] is a known
indicator of intrinsic AGN power, these correlations confirm AGN
illumination of the circum-nuclear dust as the primary heating
mechanism for the dust producing thermal MFIR emission at both 24 and
70\moo. We demonstrate that AGN heating is energetically feasible, and
identify the narrow line region clouds as the most likely location of
the cool, far-IR emitting dust. Starbursts make a major
contribution to the heating of the cool dust in only 15-28\% of our
targets.

 We also investigate the orientation dependence of the continuum
properties, finding that the broad- and narrow-line objects in our
sample with strong emission lines have similar distributions of MFIR
luminosities and colours. Therefore our results are entirely
consistent with the orientation-based unified schemes for powerful
radio galaxies. However, the weak line radio galaxies (WLRG) form a
separate class of objects with intrinsically low luminosity AGN in
which both the optical emission lines and the MFIR continuum are weak.

\end{abstract}


\section{Introduction}
\label{sec:intro}
In the last few decades space-based infrared observatories have
provided new opportunities to understand the physics of AGN. In
particular, mid- to far-infrared (MFIR) observations can address
many outstanding issues relating to the origin of the prodigious
emission, as well as bringing insight to questions relating to the
unification of active galaxies, the triggering of the activity, and
the evolution of AGN in general.

Several efforts have been made to model the MFIR spectral energy
distributions (SEDs) of radio-loud AGN in order to understand the
origin of the MFIR emission. The thermal emission from warm dust
radiating in the mid-infrared (mid-IR) is readily explained in models
by AGN heating of dust close to the active core (\citealp{pier92},
\citealp{pier93}, \citealp{granato94}), but the origin of the thermal
far-infrared (far-IR) emission is less well understood. This is
because the far-IR continuum is emitted by cooler dust ($<50$K), and
models that attempt to explain the heating of the cool dust in terms
of AGN illumination of a compact optically thick dust torus have
struggled to explain how radiation from the AGN can heat dust at
sufficiently large radii to produce emission in the
far-IR. Alternative theoretical studies that focus on adjusting the
geometry and size of the torus, or modelling it as a collection of
discrete dust clouds surrounding the AGN (e.g. \citealp{nenkova02},
\citealp{vanbemmel03}, \citealp{fritz06}), can account for the far-IR
emission in terms of AGN illumination. However, the main problem with
testing such models is our lack of direct knowledge of the spatial
distribution of the emitting dust. Due to their large distances, the
circum-nuclear dust structures remain unresolved for the vast majority
of AGN. Without constraints on the radial distribution of the dust,
there is a limit to how much the SED modelling approach can inform us
about the origin of the MFIR emission.

 Statistical studies that correlate MFIR, optical and radio properties
provide a promising alternative to direct, spatially resolved, studies
of the dust. This is due to the fact that the MFIR continuum is likely
to be more isotropic than the shorter wavelength UV-optical-near
infrared continuum, and also because the thermal emission from
circum-nuclear dust acts as a bolometer for the AGN activity. MFIR
observations of radio-loud AGN are particularly important
because the extended radio lobe emission from such galaxies is
generally considered to be emitted isotropically, providing an
opportunity to select orientation-unbiased samples.

Many previous studies have acknowledged these benefits
(\citealp{golombek88}, \citealp{impey93},
\citealp{heckman92,heckman94}, \citealp{hes95}, \citealp{haas04},
\citealp{shi05}, \citealp{cleary07}). However, definitive results were
hampered in the past by the relatively low sensitivity of the infrared
observatories, as well as the lack of complete, homogeneous samples of
powerful radio galaxies in the local Universe. It is notable that IRAS
and ISO detected less than 30\% and 50\% respectively of powerful 3C
radio galaxies at moderate redshifts ($z<0.3$). IRAS-based studies
such as those by \citet{heckman94} and \citet{hes95}, established
putative correlations between low frequency radio and MFIR emission
for radio-loud AGN, indicating a link between the AGN power and the
MFIR continuum emission. However, due to the limited detection rate in
the MFIR, the heating mechanism for the dust producing the prodigious
MFIR emission remained uncertain.

It is now widely accepted that the mid-IR ($<$30\moo) continuum is
heated by direct AGN illumination of dust structures close to the AGN
(e.g. the circum-nuclear torus). However, in light of the large
scatter of correlations between optical and far-IR ($>30\mu \rm{m}$)
continuum properties, it has been suggested that illumination by a
starburst component may provide the principal heating mechanism for
the dust producing the far-IR emission (\citealp{rowan95}). More
direct observational evidence for starburst heating of the far-IR
continuum in AGN is presented in \citet{schweitzer06} and
\citet{netzer07}, based on an analysis of starburst-sensitive
polycyclic aromatic hydrocarbon (PAH) features. Finding a correlation
between PAH 7.7\moo\ and 60\moo\ luminosity in a sample of nearby PG
quasars, \citet{schweitzer06} and \citet{netzer07} attribute this to a
link between far-IR luminosity and star formation. However, their
sample is modest, and the PAH star formation signature remains
undetected in 60\% of the objects. Therefore the putative correlation
between PAH and far-IR continuum properties lacks a solid statistical
foundation. While there is no doubt that starburst heating of cool
dust can account for a substantial fraction of the far-IR flux in
\emph{some} AGN, the starburst contribution to the far-IR continuum in
the general population of AGN remains uncertain.

As well as their importance for understanding the main dust heating
mechanism, MFIR observations can also be used to test the
orientation-based unified schemes \citep{barthel89}, under the
assumption that the MFIR emission is isotropic. To date, the results
from such tests have been ambiguous. Early IRAS-based studies
(\citealp{heckman94}, \citealp{hes95}) presented evidence for stronger
MFIR emission in broad line radio galaxies and quasars (BLRG/Q)
compared with narrow line radio galaxies (NLRG), suggesting that the
MFIR is not, in fact, isotropic. On the other hand, additional studies
using ISO data (\citealp{meisenheimer01}, \citealp{haas04}) found no
evidence for differences between the MFIR luminosities of the two
optical classes. Unfortunately, these studies were hampered by the
relatively poor sensitivity of the IRAS and ISO observatories. Further
studies using Spitzer found evidence for a difference between the MFIR
emission of BLRG/Q and NLRG. \citet{shi05} attribute this difference
to anisotropic emission at mid-IR wavelenghts, while \citet{cleary07}
attribute it to a combination of non-thermal contamination of the MFIR
emission as well as anisotropic emission at mid-IR
wavelenghts. However, both these results were based on samples that
were heterogeneous and/or incomplete in terms of far-IR detections.

Many of the past studies of MFIR emission from radio galaxies have
selected samples based on the 3C radio catalogue. Exploitation of this
catalogue is at present hampered by the lack of published high quality
optical spectroscopic observations for many of the objects, with which
one can confidently classify and identify possible links
between MFIR emission, starburst and AGN activity. In contrast, the
southern 2Jy sample \citep{tadhunter93} is unique in the sense that
deep spectra have been published for the whole sample
(\citealp{tadhunter93,tadhunter98,tadhunter02}; \citealp{wills04},
\citealp{holt07}: See \S \ref{sec:sample}). The completeness and
availability of deep spectroscopic and radio data make this sample
well suited to investigating the nature of the MFIR emission and
testing the unified schemes. Therefore we have undertaken a program of
deep imaging with Spitzer/MIPS of the 2Jy sample, in order to address
the sensitivity problems of previous MFIR observatories leading
to incomplete sample statistics (see \citealp{dicken08}, hereafter
D08).

A preliminary analysis of the Spitzer/MIPS 2Jy data set was presented
in \cite{tadhunter07} and the measured Spitzer MFIR fluxes are
presented in D08. In this paper we conduct an in-depth analysis of the
results: sections 2 and 3 present the sample and the data; section 4
is concerned with the origin of the MFIR emission; section 5 analyses
the far-IR emission and the contribution of starburst heating to the
emission at these wavelengths; and section 6 discusses these results
in the context of the heating mechanism, the covering factor of the
MFIR-emitting dust, the slopes of the correlations and the unified
schemes for powerful radio galaxies.

\section{The sample}
\label{sec:sample}
The sample selected for this study comprises a complete sample of all
46 powerful radio galaxies and steep-spectrum quasars ($ F_{\nu}
\propto \nu^{-\alpha},\alpha^{4.8}_{2.7} > 0.5 $)\footnote{In addition
to excluding quasars with $\alpha^{4.8}_{2.7} > 0.5 $, we also
excluded the quasars PKS0159-11 and PKS0842-75 on the basis that they
have relatively strong unresolved radio core emission that pushes them
above the 2Jy flux limit for the sample as a whole.} selected from the
2Jy sample of \citet{wall85} with redshifts 0.05 $<$ z $<$ 0.7, flux
densities $S_{2.7\rm{GHz}}>$ 2Jy and declinations $\delta<10^o$. This
sample is a complete, redshift limited, sub-set of that presented in
\citet{tadhunter93}, with the addition of \object{PKS 0347$+$05},
which has since proved to fulfil the same selection criteria
\citep{diserego94}. Our selection criteria cut out all the 16
quasar-like objects in the full 2Jy sample that only meet the 2Jy flux
criterion because of the strength of their beamed flat spectrum radio
core/inner jet components\footnote{The only potentially ambiguous
cases are PKS0521-36 and 3C273, which have flat spectra based on their
integrated radio emission, but extended steep spectrum emission
components with flux $S_{2.7GHz}>2$Jy. However, given that the
extended steep spectrum emission in these objects is concentrated in
one-sided jets that are likely to be strongly beamed and hence likely
to be dominated by non-thermal emission, we have decided not to
include them in the analysis presented in this paper. We find that the
inclusion/exclusion of these objects in the statistical tests makes no
difference to the main conclusions of the paper.}; the remaining
objects in the sample are all dominated by their extended steep
spectrum lobe/hotspot emission. Therefore, given that there is no
strong evidence for beaming and anisotropy in extended steep spectrum
radio components, our sample is unlikely to be significantly biased
towards a particular orientation of the jets to the line of sight. The
lower redshift limit has been set to ensure that these galaxies are
genuinely powerful sources. Further discussion of the sample selection
can be found in D08.

In the detailed analysis of the spectral energy distributions presented
in D08, we showed that a maximum of 30\% of our complete sample have
the possibility of contamination of their MFIR flux by non-thermal
synchrotron emission. This is consistent with several previous studies
of samples of radio sources (\citealp{polletta00},
\citealp{cleary07}), which have indicated that the number of
objects with possible non-thermal contamination of the MFIR is small, and
generally confined to quasars with flat spectrum radio cores.

\clearpage
\begin{deluxetable}{c@{\hspace{0mm}}l@{\hspace{-2mm}}c@{\hspace{-3mm}}c@{\hspace{-1mm}}c@{\hspace{-0mm}}c@{\hspace{-3mm}}r@{\hspace{0mm}}r@{\hspace{0mm}}c@{\hspace{0mm}}c}
\tabletypesize{\scriptsize}
\tablecaption{The Sample \label{tbl-1}}
\tablewidth{0pt}
\tablehead{
\colhead{PKS}{\hspace{0mm}} & \colhead{Other}{\hspace{-2mm}} &\colhead{Optical}{\hspace{-3mm}} & \colhead{Radio}{\hspace{-1mm}} & \colhead{z}{\hspace{0mm}}& \colhead{$L_{24}$(W/Hz)}{\hspace{-2mm}} & \colhead{$L_{70}$(W/Hz)}{\hspace{-2mm}} & \colhead{$L_{[\rm{OIII}]}$(W)}{\hspace{-3mm}} & \colhead{$L_{radio}^{5GHz}$(W/Hz)} & \colhead{Starburst}
}
\startdata
0023$-$26	&	\phantom{aa}		&	\phantom{a}	NLRG	&	CSS	&	\phantom{a}	0.322	&$	\phantom{a}	4.6\times10^{23	\phantom{a}	}$&$	5.6\times10^{24	\phantom{aa}	}$&$	6.5\times10^{34	}$&$	4.7\times10^{26	}$&	\phantom{aa}	SB	\\
0034$-$01	&	\phantom{aa}	3C015	&	\phantom{a}	WLRG	&	FRII	&	\phantom{a}	0.073	&$	\phantom{a}	4.2\times10^{22	\phantom{a}	}$&$	1.0\times10^{23	\phantom{aa}	}$&$	1.4\times10^{33	}$&$	9.0\times10^{24	}$&	\phantom{aa}	No	\\
0035$-$02	&	\phantom{aa}	3C17	&	\phantom{a}	BLRG	&	(FRII)	&	\phantom{a}	0.220	&$	\phantom{a}	7.1\times10^{23	\phantom{a}	}$&$	1.4\times10^{24	\phantom{aa}	}$&$	5.4\times10^{34	}$&$	7.6\times10^{25	}$&	\phantom{aa}	No	\\
0038$+$09	&	\phantom{aa}	3C18	&	\phantom{a}	BLRG	&	FRII	&	\phantom{a}	0.188	&$	\phantom{a}	9.9\times10^{23	\phantom{a}	}$&$	1.2\times10^{24	\phantom{aa}	}$&$	6.8\times10^{34	}$&$	7.3\times10^{25	}$&	\phantom{aa}	No	\\
0039$-$44	&	\phantom{aa}		&	\phantom{a}	NLRG	&	FRII	&	\phantom{a}	0.346	&$	\phantom{a}	5.3\times10^{24	\phantom{a}	}$&$	1.1\times10^{25	\phantom{aa}	}$&$	5.1\times10^{35	}$&$	2.0\times10^{26	}$&	\phantom{aa}	No	\\
0043$-$42	&	\phantom{aa}		&	\phantom{a}	WLRG	&	FRII	&	\phantom{a}	0.116	&$	\phantom{a}	1.5\times10^{23	\phantom{a}	}$&$	1.3\times10^{23	\phantom{aa}	}$&$	2.2\times10^{33	}$&$	4.2\times10^{25	}$&	\phantom{aa}	No	\\
0105$-$16	&	\phantom{aa}	3C32	&	\phantom{a}	NLRG	&	FRII	&	\phantom{a}	0.400	&$	\phantom{a}	1.8\times10^{24	\phantom{a}	}$&$<	2.2\times10^{24	\phantom{aa}	}$&$	1.1\times10^{35	}$&$	3.0\times10^{26	}$&	\phantom{aa}	No	\\
0117$-$15	&	\phantom{aa}	3C38	&	\phantom{a}	NLRG	&	FRII	&	\phantom{a}	0.565	&$	\phantom{a}	3.7\times10^{24	\phantom{a}	}$&$	1.2\times10^{25	\phantom{aa}	}$&$	7.2\times10^{35	}$&$	7.3\times10^{26	}$&	\phantom{aa}	No	\\
0213$-$13	&	\phantom{aa}	3C62	&	\phantom{a}	NLRG	&	FRII	&	\phantom{a}	0.147	&$	\phantom{a}	8.9\times10^{23	\phantom{a}	}$&$	8.2\times10^{23	\phantom{aa}	}$&$	5.6\times10^{34	}$&$	4.4\times10^{25	}$&	\phantom{aa}	No	\\
0235$-$19	&	\phantom{aa}	OD-159	&	\phantom{a}	BLRG	&	FRII	&	\phantom{a}	0.620	&$	\phantom{a}	5.5\times10^{24	\phantom{a}	}$&$	7.1\times10^{24	\phantom{aa}	}$&$	8.6\times10^{35	}$&$	9.8\times10^{26	}$&	\phantom{aa}	No	\\
0252$-$71	&	\phantom{aa}		&	\phantom{a}	NLRG	&	CSS	&	\phantom{a}	0.566	&$	\phantom{a}	1.7\times10^{24	\phantom{a}	}$&$<	5.5\times10^{24	\phantom{aa}	}$&$	6.4\times10^{34	}$&$	9.7\times10^{26	}$&	\phantom{aa}	No	\\
0347$+$05	&	\phantom{aa}		&	\phantom{a}	BLRG	&	FRII	&	\phantom{a}	0.339	&$	\phantom{a}	8.0\times10^{23	\phantom{a}	}$&$	7.1\times10^{24	\phantom{aa}	}$&$	9.1\times10^{33	}$&$	2.0\times10^{26	}$&	\phantom{aa}	U	\\
0349$-$27	&	\phantom{aa}		&	\phantom{a}	NLRG	&	FRII	&	\phantom{a}	0.066	&$	\phantom{a}	4.2\times10^{22	\phantom{a}	}$&$	2.0\times10^{23	\phantom{aa}	}$&$	5.3\times10^{33	}$&$	9.2\times10^{24	}$&	\phantom{aa}	U	\\
0404$+$03	&	\phantom{aa}	3C105	&	\phantom{a}	NLRG	&	FRII	&	\phantom{a}	0.089	&$	\phantom{a}	2.6\times10^{23	\phantom{a}	}$&$	6.0\times10^{23	\phantom{aa}	}$&$	1.3\times10^{34	}$&$	2.0\times10^{25	}$&	\phantom{aa}	No	\\
0409$-$75	&	\phantom{aa}		&	\phantom{a}	NLRG	&	FRII	&	\phantom{a}	0.693	&$	\phantom{a}	2.3\times10^{24	\phantom{a}	}$&$	1.7\times10^{25	\phantom{aa}	}$&$	5.6\times10^{34	}$&$	3.6\times10^{27	}$&	\phantom{aa}	SB	\\
0442$-$28	&	\phantom{aa}		&	\phantom{a}	NLRG	&	FRII	&	\phantom{a}	0.147	&$	\phantom{a}	5.1\times10^{23	\phantom{a}	}$&$	7.2\times10^{23	\phantom{aa}	}$&$	3.1\times10^{34	}$&$	5.6\times10^{25	}$&	\phantom{aa}	No	\\
0620$-$52	&	\phantom{aa}		&	\phantom{a}	WLRG	&	FRI	&	\phantom{a}	0.051	&$	\phantom{a}	1.3\times10^{22	\phantom{a}	}$&$	1.4\times10^{23	\phantom{aa}	}$&$<	1.1\times10^{32	}$&$	3.3\times10^{24	}$&	\phantom{aa}	SB	\\
0625$-$35	&	\phantom{aa}	OH-342	&	\phantom{a}	WLRG	&	FRI	&	\phantom{a}	0.055	&$	\phantom{a}	7.6\times10^{22	\phantom{a}	}$&$	1.4\times10^{23	\phantom{aa}	}$&$	1.3\times10^{33	}$&$	6.5\times10^{24	}$&	\phantom{aa}	No	\\
0625$-$53	&	\phantom{aa}		&	\phantom{a}	WLRG	&	FRII	&	\phantom{a}	0.054	&$	\phantom{a}	5.5\times10^{21	\phantom{a}	}$&$<	3.4\times10^{22	\phantom{aa}	}$&$<	4.9\times10^{32	}$&$	4.7\times10^{24	}$&	\phantom{aa}	No	\\
0806$-$10	&	\phantom{aa}	3C195	&	\phantom{a}	NLRG	&	FRII	&	\phantom{a}	0.110	&$	\phantom{a}	3.4\times10^{24	\phantom{a}	}$&$	6.4\times10^{24	\phantom{aa}	}$&$	2.6\times10^{35	}$&$	2.2\times10^{25	}$&	\phantom{aa}	No	\\
0859$-$25	&	\phantom{aa}		&	\phantom{a}	NLRG	&	FRII	&	\phantom{a}	0.305	&$	\phantom{a}	9.2\times10^{23	\phantom{a}	}$&$	8.3\times10^{23	\phantom{aa}	}$&$	4.2\times10^{34	}$&$	2.4\times10^{26	}$&	\phantom{aa}	No	\\
0915$-$11	&	\phantom{aa}	Hydra A	&	\phantom{a}	WLRG	&	FRI	&	\phantom{a}	0.054	&$	\phantom{a}	2.9\times10^{22	\phantom{a}	}$&$	3.8\times10^{23	\phantom{aa}	}$&$	1.3\times10^{33	}$&$	4.2\times10^{25	}$&	\phantom{aa}	SB	\\
0945$+$07	&	\phantom{aa}	3C227	&	\phantom{a}	BLRG	&	FRII	&	\phantom{a}	0.086	&$	\phantom{a}	3.3\times10^{23	\phantom{a}	}$&$	1.3\times10^{23	\phantom{aa}	}$&$	3.5\times10^{34	}$&$	2.1\times10^{25	}$&	\phantom{aa}	U	\\
1136$-$13	&	\phantom{aa}		&	\phantom{a}	Q	&	FRII	&	\phantom{a}	0.554	&$	\phantom{a}	6.1\times10^{24	\phantom{a}	}$&$	1.1\times10^{25	\phantom{aa}	}$&$	2.4\times10^{36	}$&$	8.9\times10^{26	}$&	\phantom{aa}	U	\\
1151$-$34	&	\phantom{aa}		&	\phantom{a}	Q	&	CSS	&	\phantom{a}	0.258	&$	\phantom{a}	1.5\times10^{24	\phantom{a}	}$&$	4.7\times10^{24	\phantom{aa}	}$&$	1.2\times10^{35	}$&$	2.3\times10^{26	}$&	\phantom{aa}	U	\\
1306$-$09	&	\phantom{aa}		&	\phantom{a}	NLRG	&	CSS	&	\phantom{a}	0.464	&$	\phantom{a}	2.0\times10^{24	\phantom{a}	}$&$	9.1\times10^{24	\phantom{aa}	}$&$	6.3\times10^{34	}$&$	5.9\times10^{26	}$&	\phantom{aa}	U	\\
1355$-$41	&	\phantom{aa}		&	\phantom{a}	Q	&	FRII	&	\phantom{a}	0.313	&$	\phantom{a}	6.0\times10^{24	\phantom{a}	}$&$	7.5\times10^{24	\phantom{aa}	}$&$	3.4\times10^{35	}$&$	2.0\times10^{26	}$&	\phantom{aa}	U	\\
1547$-$79	&	\phantom{aa}		&	\phantom{a}	BLRG	&	FRII	&	\phantom{a}	0.483	&$	\phantom{a}	2.9\times10^{24	\phantom{a}	}$&$	7.0\times10^{24	\phantom{aa}	}$&$	1.2\times10^{36	}$&$	5.1\times10^{26	}$&	\phantom{aa}	U	\\
1559$+$02	&	\phantom{aa}	3C327	&	\phantom{a}	NLRG	&	FRII	&	\phantom{a}	0.104	&$	\phantom{a}	2.8\times10^{24	\phantom{a}	}$&$	5.5\times10^{24	\phantom{aa}	}$&$	7.8\times10^{34	}$&$	3.4\times10^{25	}$&	\phantom{aa}	No	\\
1602$+$01	&	\phantom{aa}	3C327.1	&	\phantom{a}	BLRG	&	FRII	&	\phantom{a}	0.462	&$	\phantom{a}	2.2\times10^{24	\phantom{a}	}$&$	3.5\times10^{24	\phantom{aa}	}$&$	2.9\times10^{35	}$&$	3.4\times10^{26	}$&	\phantom{aa}	No	\\
1648$+$05	&	\phantom{aa}	Herc A	&	\phantom{a}	WLRG	&	FRI	&	\phantom{a}	0.154	&$	\phantom{a}	6.6\times10^{22	\phantom{a}	}$&$<	6.2\times10^{23	\phantom{aa}	}$&$	2.0\times10^{33	}$&$	3.7\times10^{26	}$&	\phantom{aa}	No	\\
1733$-$56	&	\phantom{aa}		&	\phantom{a}	BLRG	&	FRII	&	\phantom{a}	0.098	&$	\phantom{a}	3.3\times10^{23	\phantom{a}	}$&$	1.7\times10^{24	\phantom{aa}	}$&$	2.9\times10^{34	}$&$	3.5\times10^{25	}$&	\phantom{aa}	U	\\
1814$-$63	&	\phantom{aa}		&	\phantom{a}	NLRG	&	CSS	&	\phantom{a}	0.063	&$	\phantom{a}	2.5\times10^{23	\phantom{a}	}$&$	5.9\times10^{23	\phantom{aa}	}$&$	1.9\times10^{33	}$&$	1.4\times10^{25	}$&	\phantom{aa}	U	\\
1839$-$48	&	\phantom{aa}		&	\phantom{a}	WLRG	&	FRI	&	\phantom{a}	0.112	&$	\phantom{a}	4.5\times10^{22	\phantom{a}	}$&$	1.6\times10^{23	\phantom{aa}	}$&$<	1.0\times10^{32	}$&$	1.8\times10^{25	}$&	\phantom{aa}	No	\\
1932$-$46	&	\phantom{aa}		&	\phantom{a}	BLRG	&	FRII	&	\phantom{a}	0.231	&$	\phantom{a}	2.1\times10^{23	\phantom{a}	}$&$	1.5\times10^{24	\phantom{aa}	}$&$	1.1\times10^{35	}$&$	2.4\times10^{26	}$&	\phantom{aa}	SB	\\
1934$-$63	&	\phantom{aa}		&	\phantom{a}	NLRG	&	CSS	&	\phantom{a}	0.183	&$	\phantom{a}	6.2\times10^{23	\phantom{a}	}$&$	7.1\times10^{23	\phantom{aa}	}$&$	5.2\times10^{34	}$&$	2.6\times10^{26	}$&	\phantom{aa}	No	\\
1938$-$15	&	\phantom{aa}		&	\phantom{a}	BLRG	&	FRII	&	\phantom{a}	0.452	&$	\phantom{a}	2.3\times10^{24	\phantom{a}	}$&$	6.6\times10^{24	\phantom{aa}	}$&$	3.4\times10^{35	}$&$	9.2\times10^{26	}$&	\phantom{aa}	No	\\
1949$+$02	&	\phantom{aa}	3C403	&	\phantom{a}	NLRG	&	FRII	&	\phantom{a}	0.059	&$	\phantom{a}	6.9\times10^{23	\phantom{a}	}$&$	1.2\times10^{24	\phantom{aa}	}$&$	3.2\times10^{34	}$&$	8.6\times10^{24	}$&	\phantom{aa}	No	\\
1954$-$55	&	\phantom{aa}		&	\phantom{a}	WLRG	&	FRI	&	\phantom{a}	0.060	&$	\phantom{a}	1.0\times10^{22	\phantom{a}	}$&$	3.3\times10^{22	\phantom{aa}	}$&$<	4.5\times10^{31	}$&$	6.1\times10^{24	}$&	\phantom{aa}	No	\\
2135$-$14	&	\phantom{aa}		&	\phantom{a}	Q	&	FRII	&	\phantom{a}	0.200	&$	\phantom{a}	4.5\times10^{24	\phantom{a}	}$&$	4.9\times10^{24	\phantom{aa}	}$&$	5.9\times10^{35	}$&$	6.5\times10^{25	}$&	\phantom{aa}	U	\\
2135$-$20	&	\phantom{aa}	OX-258	&	\phantom{a}	BLRG	&	CSS	&	\phantom{a}	0.635	&$	\phantom{a}	6.2\times10^{24	\phantom{a}	}$&$	5.3\times10^{25	\phantom{aa}	}$&$	6.4\times10^{35	}$&$	1.1\times10^{27	}$&	\phantom{aa}	SB	\\
2211$-$17	&	\phantom{aa}	3C444	&	\phantom{a}	WLRG	&	FRII	&	\phantom{a}	0.153	&$	\phantom{a}	1.8\times10^{22	\phantom{a}	}$&$<	3.4\times10^{23	\phantom{aa}	}$&$	1.1\times10^{33	}$&$	6.2\times10^{25	}$&	\phantom{aa}	No	\\
2221$-$02	&	\phantom{aa}	3C445	&	\phantom{a}	BLRG	&	FRII	&	\phantom{a}	0.057	&$	\phantom{a}	7.4\times10^{23	\phantom{a}	}$&$	5.9\times10^{23	\phantom{aa}	}$&$	7.6\times10^{34	}$&$	7.6\times10^{24	}$&	\phantom{aa}	No	\\
2250$-$41	&	\phantom{aa}		&	\phantom{a}	NLRG	&	FRII	&	\phantom{a}	0.310	&$	\phantom{a}	1.4\times10^{24	\phantom{a}	}$&$	2.7\times10^{24	\phantom{aa}	}$&$	2.2\times10^{35	}$&$	1.7\times10^{26	}$&	\phantom{aa}	No	\\
2314$+$03	&	\phantom{aa}	3C459	&	\phantom{a}	NLRG	&	FRII	&	\phantom{a}	0.220	&$	\phantom{a}	4.0\times10^{24	\phantom{a}	}$&$	4.0\times10^{25	\phantom{aa}	}$&$	7.2\times10^{34	}$&$	8.1\times10^{25	}$&	\phantom{aa}	SB	\\
2356$-$61	&	\phantom{aa}		&	\phantom{a}	NLRG	&	FRII	&	\phantom{a}	0.096	&$	\phantom{a}	4.0\times10^{23	\phantom{a}	}$&$	7.3\times10^{23	\phantom{aa}	}$&$	4.0\times10^{34	}$&$	4.7\times10^{25	}$&	\phantom{aa}	No	\\

\enddata

\tablecomments{Table\ref{tbl-1}: Column 3 definitions; Q - quasar,
  BLRG - broad line radio galaxy, NLRG - narrow line radio galaxy,
  WLRG - weak line radio galaxy. Column 4 definitions; FRI \&\ FRII -
  Fanaroff-Riley class 1 and 2 respectively, CSS - compact steep
  spectrum, C/J - core/jet. Column 8 [\rm{OIII}]$\lambda5007$
  luminosities calculated from flux presented in \citet{tadhunter93}
  and \citet{wills04}. Column 10 gives the 15-17Ghz monochromatic core
  luminosities from the fluxes presented in D08. Column 11 gives
  information about whether a spectroscopic young stellar population
  (YSP) has been detected at optical wavelengths indicating the
  possibility of starburst activity.  SB - YSP detected, No - No YSP, U
  - uncertain YSP component (references for YSP/starburst:
  \citealp{tadhunter02};\citealp{wills04,wills08}; \citealp{holt07}). Positions
  for the objects can be found in D08. }

\end{deluxetable}
\clearpage

\section{The data}
\label{sec:data}
We have made deep Spitzer/MIPS observations (24, 70 and 160\moo)
of our sample, as well as complementary high frequency radio
observations (15 to 22 \rm{GHz}) with the ATCA and the VLA. Full details of
the observations and reduction can be found in D08, along with the
MFIR, radio fluxes and spectral energy distributions for the entire
sample. We detect 100\% of our sample at 24\moo, 90\% at 70\moo\ and
33\% in the lower sensitivity 160\moo\ band. This is by far the best
detection rate for MFIR observations of a sample
of an intermediate-redshift AGN published to
date. Additionally, the high frequency radio data, along with data
from the literature, have enabled us to detect the radio cores in
$\approx$70\% of our complete sample. We utilize these data to
calculate the orientation-sensitive R parameter (see D08 and \S \ref{sec:R}).

In Table \ref{tbl-1} we present the 24 and 70\moo\ monochromatic
luminosities for the complete sample, along with 5 \rm{GHz} radio and
[\rm{OIII}] $\lambda5007$ emission line luminosities\footnote{In order
to calculate luminosities we used $H_{o}=71km s^{-1} Mpc^{-1},
\Omega_{m}=0.27$ and $\Omega_{\lambda}=0.73$ along with spectral
indices derived from the F(70)/F(24) flux ratios}. The analysis
presented here differs from the preliminary results presented in
\citet{tadhunter07} in that the emission line luminosities have now
been corrected for Galactic extinction, using E(B-V) reddening values
obtained from the NASA/IPAC Extragalactic Database (NED), along with
the parameterised Galactic extinction law of \citet{howarth83}. In Table
\ref{tbl-1} we also present the most up-to-date optical classification
for each of the sources, with objects classified as narrow-line radio
galaxies (NLRG), broad-line radio galaxies or quasars (BLRG/Q), and
weak-line radio galaxies (WLRG)\footnote{WLRGs are sometimes
known as low-excitation galaxies but we prefer to label them as WLRG
since the excitation (or, more accurately, ionization) of the emission
line gas is not necessarily related to the AGN luminosity. Indeed,
there exist examples of AGN with relatively high [OIII] luminosities,
but emission line ratios reflecting a low ionization state. WLRG are
defined as having EW([OIII])$<$10\AA\ \citep{tadhunter98}.}.

\begin{figure}[t]
\epsscale{1}
\plotone{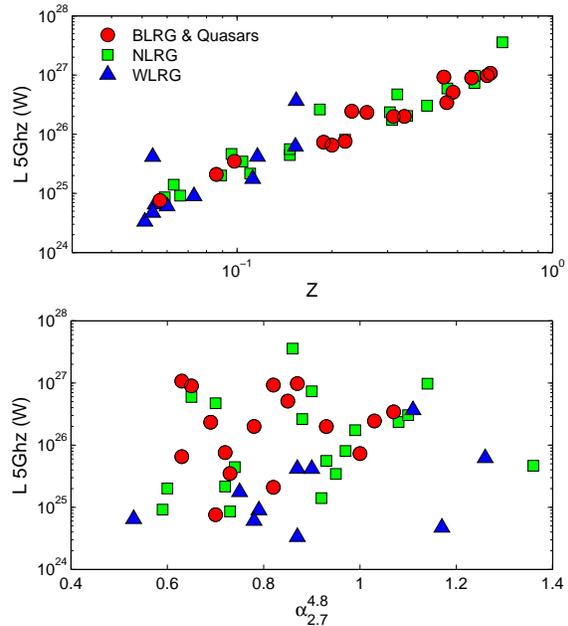}
\caption{\label{fig:rad_z} Distribution of 5 \rm{GHz} vs. redshift
total radio luminosity (top); high frequency radio spectral index
($\alpha^{4.8}_{2.7}$) vs. 5 \rm{GHz} radio luminosity (bottom) for
the 2Jy sample.}
\end{figure}

In Figure \ref{fig:rad_z} we plot 5 \rm{GHz} radio luminosity against
both redshift (top) and radio spectral index (bottom) for the complete
sample. The effect of the flux limit of the sample is clearly visible
in the upper plot in the form of the tight correlation between
monochromatic radio luminosity and redshift. It is noteworthy that our
complete sample covers three orders of magnitude in radio
luminosity. The plot of the high frequency radio spectral index
$\alpha^{4.8}_{2.7}$ against radio luminosity allows us to investigate
any bias in our steep spectrum selection method. Based on the similar
distributions of the different optical classes in $\alpha^{4.8}_{2.7}$
over the full 3 orders of magnitude covered by our sample, we are
confident that the steep spectrum selection has left us with no strong
bias towards BLRG/Q or NLRG classification.

\begin{sidewaysfigure*}
\includegraphics[width =1\columnwidth]{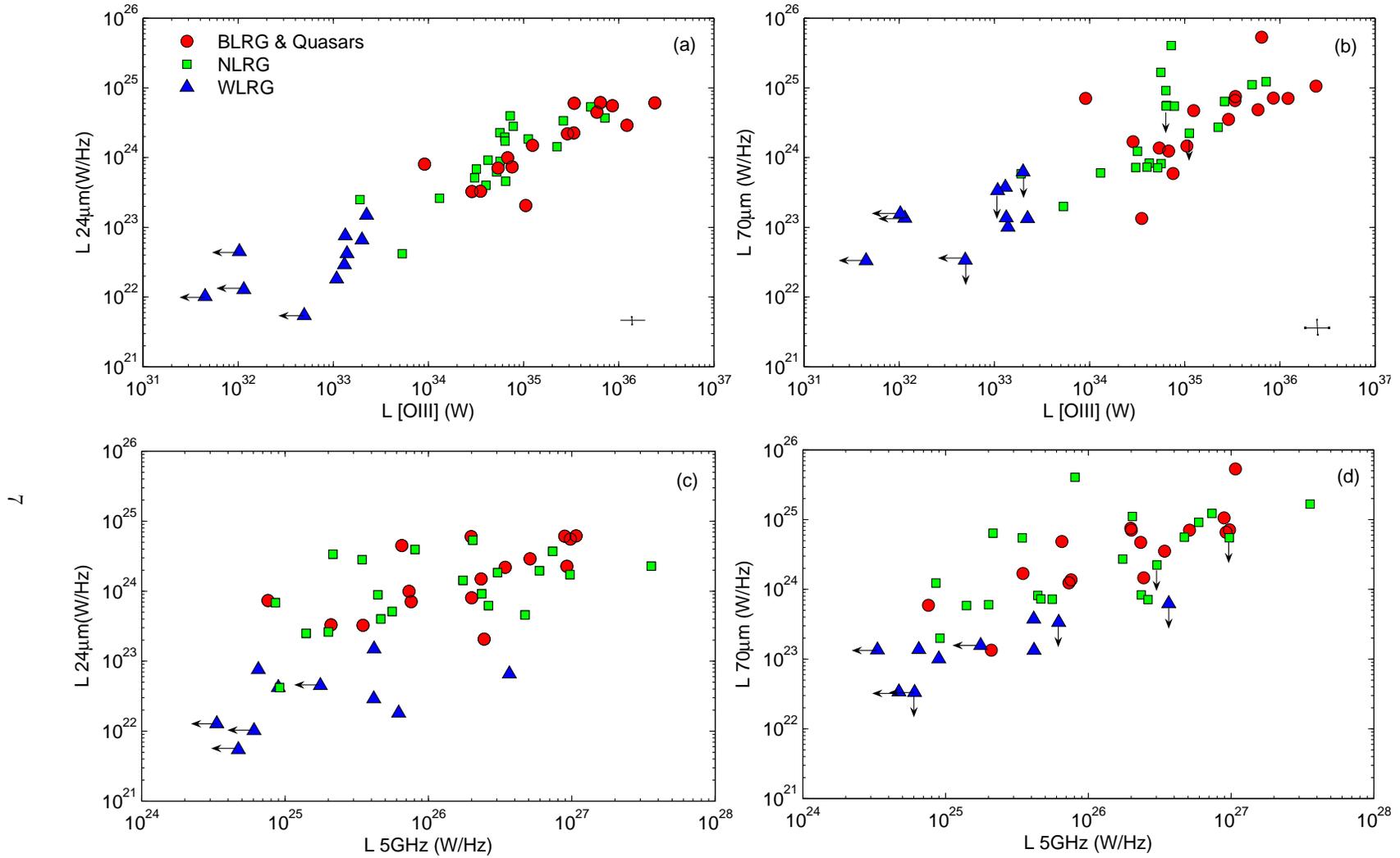}
\caption{ Luminosity correlation plots: (a) $L_{24\mu m}$
vs. $L_{[\rm{OIII}]\lambda 5007}$; (b) $L_{70\mu m}$
vs. $L_{[\rm{OIII}]\lambda 5007}$; (c) $L_{24\mu m}$
vs. $L_{5\rm{GHz}}$; (d) $L_{70\mu m}$ vs. $L_{5\rm{GHz}}$. The cross in the
bottom right corner of plots (a) and (b) represents the maximum
uncertainties in [\rm{OIII}] and MFIR luminosity measurements,
demonstrating that the scatter in the plots is real and not a
consequence of observational uncertainties. Red circles indicate
BLRG/Q objects, green squares NLRG and blue triangles
WLRG.\label{fig:opradall}}
\end{sidewaysfigure*}

\section{MFIR emission and the central engine}
\label{sec:corr}

Many previous studies (e.g. \citealp{heckman94}, \citealp{hes95},
\citealp{cleary07}) have found evidence that MFIR emission power is
correlated with measured total radio power, albeit based on highly
incomplete samples. This result is in line with a model in which the
MFIR emission and radio jet power are both strongly linked through the
physics of the central engine. However, radio emission depends on
factors in addition to the intrinsic power of the AGN. For example,
the properties of the local ISM interacting with the radio plasma are
expected to have a major impact on the conversion of jet mechanical
power into radio luminosity (e.g. \citealp{barthel96}). In addition,
the low frequency radio emission may be emitted by extended
structures that are far from the regions emitting the MFIR
continuum. Therefore, a change in the intrinsic power of the AGN may
take a significant amount of time to be reflected in the emission of
the extended radio lobes.

An alternative to comparisons based on low frequency radio emission is
to use optical emission line luminosity. The AGN-photoionized
narrow-line region (NLR) is emitted on a smaller scale ($\leq$5kpc)
than the extended radio lobe emission in most radio galaxies.
Therefore the $[\rm{OIII}]\lambda$5007 emission line is likely to
provide a good indication of the intrinsic power of the illuminating
AGN (e.g. \citealp{rawlings91}, \citealp{tadhunter98},
\citealp{simpson98}).

\subsection{Comparisons with  MFIR luminosity}
We now utilize our complete sample to address the question of how
the MFIR continuum depends on the AGN power, with the benefits of not
only completeness in the MFIR flux measurements, but also accurate
[\rm{OIII}]$\lambda$5007 emission line luminosities for the entire
sample.

In Figures \ref{fig:opradall} (a) and (b) we plot 24\moo\ and 70\moo\
monochromatic luminosities ($L_{24\mu m }$, $L_{70\mu m }$) against
[\rm{OIII}] luminosities ($L_{[\rm{OIII}]}$), while in Figures
\ref{fig:opradall} (c) and (d) we plot the 24\moo\ and 70\moo\
luminosities against 5GHz monochromatic radio luminosities
($L_{5\rm{GHz}}$). It is clear from Figures 2 (a) and 2(b) that the MFIR
luminosities are strongly correlated with $L_{[\rm{OIII}]}$. As
discussed in the preceding subsection, the [\rm{OIII}]$\lambda 5007$
luminosity is expected to provide a good indication of intrinsic AGN
power.  Therefore, we propose that the 24\moo\ and 70\moo\ emission
are also intrinsically linked to the power of the active core and,
consequently, that the dust producing the MFIR emission is likely to be
heated directly by the central AGN.

It is also notable from Figure \ref{fig:opradall}(a) that $L_{24\mu m
}$ shows a much tighter correlation with the $L_{[\rm{OIII}]}$ than it
does with the $L_{5\rm{GHz}}$. This result is expected, because we
believe that both the 24\moo\ luminosity and the [\rm{OIII}]
luminosity are strongly linked to the power of the active core via AGN
illumination of the emission line clouds and dust structures close to
the nucleus. The radio power, however, is dependent on additional
factors such as the nature of the ISM in the halo of the host galaxy
into which the jets and lobes expand.

Considering next the differences between the 24\moo\ and 70\moo\
correlations, it is apparent that there is more scatter in the
$L_{70\mu m}$ vs $L_{[\rm{OIII}]}$ correlation. $L_{70\mu m}$ also
remains better correlated with $L_{[\rm{OIII}]}$ than $L_{5\rm{GHz}}$,
but the difference is not as evident as at 24\moo. Our interpretation
of the cause of this additional scatter at 70\moo\ lies in the origin
of the far-IR emission, as we will discuss in detail in \S
\ref{sec:origin}. However, it is noteworthy that this scatter cannot
be due to the reduced sensitivity of Spitzer at longer far-IR
wavelengths because our observational errors are small compared with
the overall scatter in the distribution (see error bars in plots (a)
and (b)). Therefore, there must be a physical cause for the increased
scatter in this plot.

\subsection{Rank correlation statistics}
\label{sec:rankstat}

\begin{deluxetable}{l@{\hspace{0mm}}c}
\tabletypesize{\scriptsize}
\tablecaption{Statistical Analysis \label{tbl-2}} \tablewidth{0pt}
\tablehead{ \colhead{Rank Correlation}{\hspace{0mm}} &
\colhead{$r_s$}{\hspace{0mm}} }
\startdata 
(1) $L_{24}$ vs $L_{[\rm{OIII}]}$   	&	0.88	\\
(2) $L_{70}$ vs $L_{[\rm{OIII}]}$   	&	0.76	\\
(3) $L_{24}$ vs $L_{5GHz}$	&	0.54	\\
(4) $L_{70}$ vs $L_{5GHz}$	&	0.63	\\
(5) $L_{5\rm{GHz}}$ vs z	        &	0.93	\\
(6) $L_{5\rm{GHz}}$ vs $L_{[\rm{OIII}]}$ 	&	0.57	\\
\cutinhead{Partial Rank Correlation with z}			
(7) $L_{24}$ vs  $L_{[\rm{OIII}]}$ 	&	0.77	\\
(8) $L_{70}$ vs  $L_{[\rm{OIII}]}$  	&	0.50	\\

\enddata \tablecomments{Table \ref{tbl-2}: Result of various Spearman
rank correlation statistics. Values of
$0<r_s<1$ are given for each test, where a value close to 1 is highly
significant. This test was undertaken with a z limited sample
z$>$0.06, to remove most of the objects with upper limits in
[OIII]. In addition the object PKS1839-48 which has upper limits in
[OIII] was also removed, leaving 38 objects from our complete sample. See \S \ref{sec:rankstat} for discussion of
the 70\moo\ upper limits.}
\end{deluxetable}

Prior to interpreting our results in full, it is important to
investigate the significance of the correlations in Figure 2. We have
therefore calculated the Spearman rank correlation coefficient for the
four correlations. The test was undertaken using a redshift limited
sub-sample with z$>$0.06 in order to remove the majority of WLRGs with
upper limits in [OIII]. In addition the remaining object with an upper
limit in [OIII] (PKS1839-48) was also excluded\footnote{The exclusion
or inclusion of the [OIII] upper limit value for this object has been
investigated, and we find that it makes no significant difference to
the statistical results presented in Table \ref{tbl-2} }. In order to
quantify the effects that the four remaining upper limits in 70\moo\
have on this test, we chose to handle the correlations that included
70\moo\ data in the following way: we created replacements for each of
the four 70\moo\ upper limits by randomly selecting a 70\moo/24\moo\
flux ratio value from the distribution of measured 70\moo/24\moo\ flux
ratio values for the sample, and then multiplying this by the measured
24\moo\ flux to create a new 70\moo\ flux estimate. The 70\moo\
estimates were then converted to luminosities and included in the rank
correlation test. This process was repeated 1000 times, and the median
of the correlation coefficient for those cycles is presented in Table
\ref{tbl-2} for correlations involving 70\moo\ luminosities (i.e.rows 2,
4 and 8).

  The results show a high level of significance for all the
correlations using a two tailed test, in the sense that we can reject
the null hypothesis that the variables are uncorrelated at a $>$99.9\%
level of significance. The rank correlation statistics, presented in
Table \ref{tbl-2}, clearly show that the $L_{[\rm{OIII}]}$
vs. $L_{24\mu m}$ and $L_{[\rm{OIII}]}$ vs. $L_{70\mu m}$ correlations
are more significant than those between the radio luminosity and MFIR
and [OIII] luminosities.

For a flux-limited sample such as that considered here, it is natural
to consider the possibility that the correlations between MFIR, radio
and [OIII] luminosities might not be intrinsic, but rather arise
through their mutual dependence on redshift. For example, given the
strong correlation between radio power and redshift induced by the
radio flux limit (see Figure 1), and also the correlation between
emission line luminosity and radio power, it is possible that a
correlation between the $L_{[\rm{OIII}]}$ and the MFIR emission
luminosity might arise because the MFIR luminosities are independently
correlated with z (e.g. because of genuine redshift evolution rather
than being intrinsically correlated with $L_{[\rm{OIII}]}$). The
second part of Table \ref{tbl-2} (rows 7 and 8), shows the results of
a partial rank correlation test carried out in order to investigate
whether the correlations could be a result of a dependence on the
third variable z. In both cases we still find that the null hypothesis
that the variables are unrelated can be rejected at a $>$99.5\% level
of significance. This demonstrates that both $L_{24\mu m }$ and
$L_{70\mu m}$ are intrinsically correlated with $L_{[\rm{OIII}]}$.

\subsection{LIRG and ULIRG comparison}
\label{sec:ulirg}

\begin{figure}[t]
\epsscale{1}
\plotone{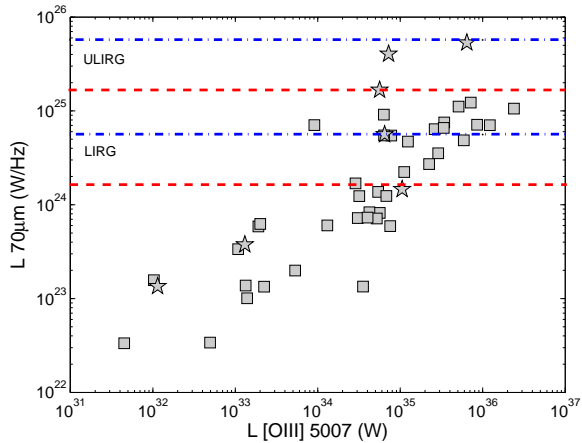}
\caption{\label{fig:ulirg} Plot of $L_{70\mu m}$ vs $L_{[OIII]}$
showing the comparison with LIRG and ULIRG luminosities. Approximate
ranges for LIRGs and ULIRGs are indicated by the horizontal lines,
where the dashed red lines, represent the luminosity limits of typical LIRGs and
ULIRGS for a warm SED model, and the dot dashed blue lines represent
the limits of LIRGs and ULIRGs for the cool SED model (see \S
\ref{sec:ulirg}). Stars represent objects with optical evidence
for recent star formation activity (see \S \ref{sec:looking}).}
\end{figure}

In order to place radio galaxies in the context of other sources of
prodigious MFIR emission in the local universe, it is interesting to
compare them with the luminous and ultra-luminous infrared galaxies
(LIRGs and ULRIGs: \citealp{sanders96}). Figure \ref{fig:ulirg} again
shows the $L_{70\mu m}$ vs $L_{[\rm{OIII}]}$ correlation for the
entire sample, but in this case we also indicate the typical ranges
that LIRGs and ULIRGs would occupy on the diagram. To calculate these
ranges we follow the definitions of \citet{sanders96}: that LIRGs have
integrated far-IR luminosities
$10^{11}L_{\odot}>L_{FIR}>10^{12}L_{\odot}$, while ULIRGs have
$10^{12}L_{\odot}>L_{FIR}>10^{13}L_{\odot}$. The ranges of 70\moo\
monochromatic luminosities corresponding to these integrated
luminosities have been estimated based on two assumptions about the
MFIR SEDs of the sources. In each case the lower line has been
calculated by assuming the SED of the radio galaxy 3C327 --- taken to
be representative of a source with relatively warm MFIR colours,
whereas the upper line has been calculated by assuming the SED of the
ULIRG Arp220 --- representative of a source with cool MFIR
colours. Considering these limits we find that 28-48\% of our sample
would be classified as LIRGs, and $\leq$7\% as ULIRGS.

\subsection{Testing the unified schemes}
\label{sec:OpClass}
The MFIR luminosities can also be used to test the orientation-based
unified schemes for powerful radio galaxies (regardless of the
emission mechanism). In particular, if the distributions of the
observed MFIR luminosities are similar for different optical classes
of objects (NLRG, BLRG/Q, WLRG) this is consistent with, but does not
prove, that they are part of the same parent population.

Evidence from investigations using IRAS (e.g. \citealp{heckman94},
\citealp{hes95}) suggested that the MFIR luminosities of narrow line
radio galaxies (NLRG) are lower by up to an order of magnitude
in comparison to broad line radio galaxies and quasars (BLRG/Q)\footnote{In
the following, unless otherwise specified, we consider BLRG and
quasars as a single class. This is justified on the basis that, among
objects with broad permitted emission lines, there is a continuous range in
luminosities between objects classed as BLRG and quasars.}. These
results have been supported by some studies (\citealt{vanbemmel00})
and rejected by others (\citealt{meisenheimer01}, \citealt{haas04}),
but all the previous studies were based on incomplete
samples. Therefore, it remains uncertain whether genuine differences
exist between the MFIR properties of broad and narrow line radio-loud
AGN.

Any difference between the two types of radio galaxies at shorter
mid-IR wavelengths ($\lesssim$30\moo), might be explained in terms
of extinction by a dusty torus, provided that there is significant
optical depth in the torus at such wavelengths. However, it would
remain challenging to explain differences in longer wavelength
emission in terms of such obscuration. Alternatively, a difference
between the two classes of objects could be due to non-thermal beamed
emission contaminating the thermal MFIR, which is expected to be
stronger in the BLRG/Q.

Using our complete sample we can investigate
whether there are significant differences between the MFIR properties
of broad- and narrow-line objects in the 2Jy sample.

A visual inspection of Figures \ref{fig:opradall} (a) and (b) reveals
no evidence that the BLRG/Q have higher luminosities than the NLRG at
24 or 70\moo,     and we have shown that no more than 24\% of the sample have
a possibility of contamination from non-thermal beamed components (see
\citet{dicken08}). Therefore we conclude that the MFIR emission in our
sample is most likely to be emitted isotropically, at least down to an
observed wavelength of 24\moo\ (rest wavelength 14-23\moo, depending
on redshift).

\begin{figure*}[t]
\epsscale{2}
\plotone{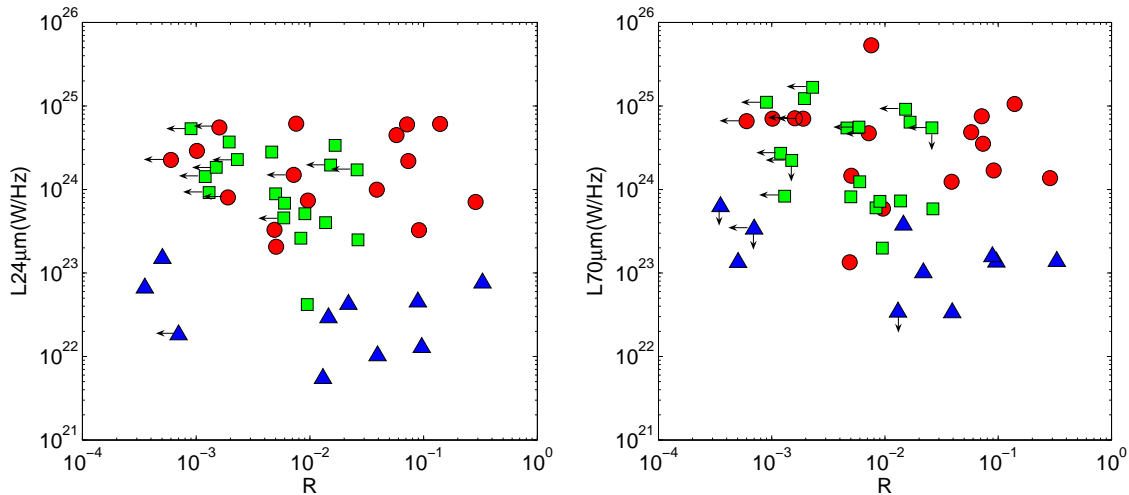}
\caption{Plot of the orientation-sensitive R vs. 24\moo\ luminosity
(left) and 70\moo\ luminosity (right), where R is defined as
$S_{core}/(S_{tot}-S_{core})$. \label{fig:R}}
\end{figure*}

\subsection{The R parameter}
\label{sec:R}
We can further examine whether the MFIR continuum is isotropic by
investigating any links between MFIR luminosities and the relative
brightness of the radio core. If there is any anisotropy in the MFIR
continuum, we would expect to see enhanced emission in those objects
with strong cores, which have axes orientated closer to the line of
sight. Using the new high frequency radio observations presented in
D08, as well as data from the literature, we can investigate the
dependence of MFIR properties on the orientation-sensitive, core
dominance parameter R, defined as
$S_{core}/(S_{tot}-S_{core})$. Tables of R values estimated at 5GHz
for the complete sample are presented in D08.

In Figure \ref{fig:R} we plot the 24 and 70\moo\ luminosities against
R. It is clear from these plots that, on average, the BLRG/Q have the
highest R values in our sample \citep{morganti97}, consistent with
unified schemes which predict that BLRG/Q have jet axes that are
pointing closer to the line of sight, leading to stronger beamed
core/jet emission. However, not all BLRG/Q are highly core
dominated. In fact, the population of such objects shows a wide range
of R parameter values.

\begin{figure}[t]
\epsscale{1}
\plotone{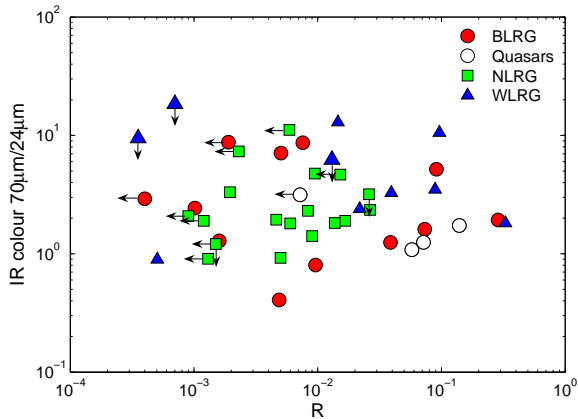}
\caption{Plot of the orientation sensitive R parameter against MFIR colour,
represented by the 70/24\moo\ flux ratio. \label{fig:colour_R}}
\end{figure}

In Figure \ref{fig:colour_R} we plot R against the 70\moo/24\moo\
infrared flux ratio. Plotting similar parameters, \citet{shi05} found
an anti-correlation in the sense that warmer IR colour corresponds to
a smaller R parameter. In addition, they found some evidence that
different optical classes lie in different regions of the plot, with
the BLRG/Q tending to the highest R, and warmest colours.

From a visual inspection of Figure \ref{fig:colour_R} there is little
evidence for a strong anti-correlation as presented in
\citet{shi05}. The range of infrared colours suggests that the MFIR
emission is relatively isotropic. However, it is worth noting that
many of the objects in the lower region of the diagram, forming the
group at low 70\moo/24\moo\ and high R, are BLRG/Q.

\subsection{Weak line radio galaxies}
It is not only the distributions of NLRG and BLRG/Q that are of
interest in the correlation plots presented in Figure
\ref{fig:opradall}. The figure also provides clues to the nature of
WLRG. Although many of the WLRG in our sample have been classified as
FRI radio sources, a significant subset have FRII morphologies (see
\citealp{tadhunter98}). Previous studies (e.g. \citealp{cao04}) have
suggested that obscuration could provide an explanation for
the differences between WLRG and radio galaxies with strong emission lines.

In Figures \ref{fig:opradall} (a) and (b) it is striking that the WLRG
lie at the lowest luminosity end in \emph{both} emission line and MFIR
continuum luminosity. If the MFIR emission is primarily caused by AGN
heating of the dust, then this demonstrates that WLRG contain
\emph{intrinsically weak} AGN. Therefore, WLRG cannot be accommodated
in the simplest unified schemes for radio galaxies, which posit that
the differences between all classes of radio-loud AGN are solely due
to anisotropy and orientation. It is evident that, although
obscuration may contribute at some level to the weakness of the
emission lines in these objects, the idea that WLRG contain powerful
quasar-like AGN, heavily obscured by dust, is certainly not consistent
with our data. \citet{hardcastle07} suggest that the differences
between WLRG and radio galaxies with strong emission lines
(NLRG/BLRG/Q) may be accounted for by different accretion modes of the
AGN, with the WLRG accreting hot ISM at a relatively low Eddington
ratio, and NLRG/BLRG/Q accreting cold ISM at a higher Eddington ratio.

Figures \ref{fig:opradall} (c) and (d) clearly show that, despite
being weak in both emission line and MFIR luminosity, the WLRG overlap
with the NLRG and BLRG/Q in terms of radio luminosity. This suggest
that other factors play an important role in boosting the radio powers
of WLRG. In particular, as noted above, the radio luminosity is not
necessarily a good measure of intrinsic AGN may power, and the
properties of the local ISM have a large impact on the radio
luminosity for a given intrinsic jet power. In this case we would
expect that, for a given MFIR luminosity, the WLRG with powerful radio
emission should be found in relatively dense cluster
environments. Some evidence for this effect is presented in
\citet{barthel96}, however, their result is based on a limited
sample. Future deep optical imaging observations of the 2Jy sample
will provide a direct indication of the significance of such
environmental effects.

\section{Origin of the Far-Infrared Emission}
\label{sec:origin}

\begin{figure*}[t]
\epsscale{2}
\plotone{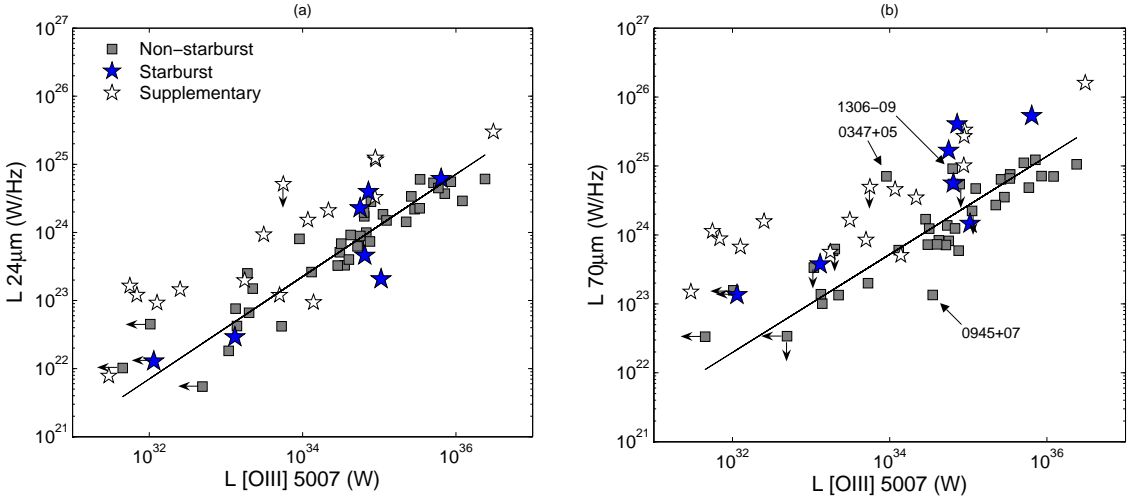}
\caption{Plots showing the variation of $L_{MFIR}$ with
$L_{\rm{[OIII]}}$ for the complete sample at 24 and 70\moo, with those
objects identified as having young stellar populations marked with
separate symbols (stars). Also displayed is a supplementary sample of
all known radio galaxies with spectroscopically-confirmed evidence for
star formation activity at optical wavelengths, the properties of
which are detailed in Table \ref{tbl-3}. The regression line has been fitted
only to the data from the original sample with z $>$ 0.06, in order to avoid
most of the objects with upper limits in their [\rm{OIII}], and also
excluding the 7 starburst objects.\label{fig:starburst}}
\end{figure*}

As discussed in \S \ref{sec:intro}, the nature of the heating
mechanism for the cool dust radiating at 70\moo\ (the far-IR) is not well
understood. However, it is widely accepted that the cool dust is
likely to be heated by starbursts and/or by AGN illumination. Based on
the tight correlation between $L_{\rm{[OIII]}}$ and $L_{MFIR}$ shown
in Figure \ref{fig:opradall}, we have concluded that the most probable
heating mechanism for the dust emitting at both mid- and far-IR
wavelengths is direct illumination by the AGN (\S
\ref{sec:corr}). However, it is of course naive to rule out starburst
heating as a contributor to far-IR emission altogether, specifically
because morphological evidence suggests that at least some powerful
radio galaxies are triggered in major, gas rich galaxy mergers
\citep{heckman86}. Such mergers are predicted to be associated with
powerful starbursts (e.g. \cite{dimatteo05}). The starburst-AGN connection is
also important for interpreting sub-millimetre observations in the
context of the star formation history of radio-loud AGN at high
redshift. Some studies, such as that by \citet{archibald01}, assume
that the cold dust responsible for the sub-mm emission is heated
entirely by starbursts. Therefore to comprehensively understand the
MFIR emission from powerful radio galaxies it is vital to investigate
the contribution of starburst heating to the observed far-IR fluxes,
and possible links between star formation and AGN activity.

\subsection{Evidence for starburst heating of the far-IR continuum}
\label{sec:looking}

Careful spectral synthesis modelling of the high quality optical
spectra for our sample (\citealp{tadhunter02}, \citealp{wills04,
wills08}, \citealp{holt07}) has allowed us to identify the objects
that show clear evidence for recent star formation activity at optical
wavelengths (see Table \ref{tbl-1}). Therefore, in Figure
\ref{fig:starburst} we are able to plot the $L_{24\mu m}$ and
$L_{70\mu m}$ data against $L_{\rm{[OIII]}}$ once again, but this time
highlighting those seven objects that we have established to have
unambiguous evidence for starburst activity at optical wavelengths
(PKS0023$-$26, PKS0409$-$75, PKS0620$-$52, PKS0915$-$11 (3C218),
PKS1932$-$46, PKS2135$-$20, PKS2314$+$03 (3C459)). As we have seen in
\S \ref{sec:corr}, there are good correlations between both $L_{24\mu
m}$, $ L_{70\mu m}$ and $L_{[\rm{OIII}]}$, but there is significantly
more scatter in the 70\moo\ correlation, consistent with the findings
of \citet{rowan95}. As first described in the preliminary results
presented in \cite{tadhunter07}, we believe this scatter is in part
due to the boosting by starburst heating of the cool far-IR emitting
dust that is not as significant for the warm mid-IR 24\moo\ dust
emission.

To demonstrate the degree of starburst boosting above the main
correlations we have plotted regression lines on both plots in Figure
\ref{fig:starburst}. The lines shown are the bisector of a linear
least squares fit of $x$ on $y$ and $y$ on $x$. As for the sample
used in the Spearman rank statistics, in calculating this fit we
include only those objects with z $>$ 0.06, in order to avoid most of
the objects with upper limits in their [\rm{OIII}] emission. In addition,
the 7 starburst objects and one more object (PKS1839$-$48) with upper
an limit in [\rm{OIII}] were also removed, leaving a total sample of
33 objects for the fit.

From a visual inspection of Figure \ref{fig:starburst}(b) it is clear
that the majority of those objects we identify as having evidence for
optical star formation activity are displaced, showing enhanced far-IR
luminosities relative to the regression line in the 70\moo\ plot; this
enhancement is not apparent for most of the optical starburst objects
in the 24\moo\ plot. Therefore, it appears that there must be some
mechanism boosting the far-IR flux in starburst compared to
non-starburst objects. Analysing the displacements of starburst
objects we find that their far-IR fluxes are boosted by a factor of
four on average relative to the remainder of the sample, with a
maximum boosting factor of 20 in the case of 3C459.

  In addition to our complete sample, we have also plotted a
supplementary sample of starburst radio galaxies, taken from the
literature on Figure \ref{fig:starburst}. These supplementary sources
represent all known radio-loud AGN that show spectroscopic evidence
for star formation activity at optical wavelengths, apart from the
similar objects in the 2Jy sample. The properties of these sources are
presented in Table \ref{tbl-3}, and they are plotted as open stars in
Figure \ref{fig:starburst}. It is clear that these supplementary
objects also lie well above the regression line fitted to our data in
the 70\moo\ plot. A visual inspection of the supplementary starburst
sample in Figure \ref{fig:starburst} also shows a tendency for
increased 24 and 70\moo\ luminosities, relative to the majority of the
sample, at low $L_{[\rm{OIII}]}$ emission. This is not surprising
given that, at low [OIII] luminosities, the AGN itself is
intrinsically weak and a modest amount of ongoing star formation can
boost the MFIR fluxes well above the regression line.

  Due to the high level of completeness of our sample in terms of MFIR
detections, our data can also be used to test the statistical
significance of any differences between the starburst and
non-starburst populations in Figure \ref{fig:starburst}.

First, for our complete sample we have considered a one dimensional
Kolmogorov-Smirnoff (K-S) two sample test, comparing the vertical
displacements from our fitted regression line in the $L_{70\mu m}$
vs. $L_{[\rm{OIII}]}$ plot (see also \citealp{tadhunter07}). In this
case we find that we can reject the null hypothesis that the starburst
and non-starburst samples are drawn from the same parent population at
a better than 1\% level of significance. We have made the same test,
this time including both the 2Jy starburst and the supplementary
starburst objects, and investigating whether they are significantly
displaced relative to the regression line for the non-starburst 2Jy
sample. In that case we find that we can reject the null hypothesis
that the starburst and non-starburst samples are drawn from the same
parent population at a level of significance of better than 0.1\%.

Secondly, we consider the significance of the displacement between the
starburst and non-starburst samples using a two dimensional K-S test
(\citealp{peacock83}, \citealp{fasano87}).  The method we apply here
is the generalisation of the 2D K-S test developed by
\citet{fasano87}. Again for the $L_{70\mu m}$ vs $L_{[\rm{OIII}]}$
plot, initially considering the starburst objects against the
non-starburst objects (in our complete 2Jy sample), we find that we
can reject the null hypothesis that the starburst and non-starburst
samples are drawn from the same parent population at a level of
significance of better than 5\%. We find exactly the same significance
level when including the supplementary starburst objects. In addition,
applying the same statistical test to the $L_{[\rm{OIII}]}$
vs. $L_{24\mu m}$ plot we find no differences between the starburst
and non-starburst samples, as expected given that the 24\moo\
luminosity is less likely to be effected by starburst heating.

There are three notable outliers from the correlation, which are
labelled on Figure \ref{fig:starburst}(b). These sources add to the
increased scatter in the far-IR, and we describe them in more detail
below.

\begin{itemize}

\item{{\bf \object{PKS0347+05}}. This object lies behind the Galactic
plane and, despite the relatively large correction of the [OIII] flux
for Galactic extinction, it still lies above our fitted regression
line. In a recent analysis of near-IR images for the source it was
found that there are, in fact, two AGN within our MFIR photometric
aperture for this object. If both AGN radiate at MFIR wavelengths this
could perhaps explain some of the apparent excess relative to the
correlation.  Additionally, since \object{PKS0347+05} is a BLRG, its
AGN is likely to outshine any starburst signatures at optical
wavelengths. Therefore, it is possible that this object could also
have star formation activity that has not so far been detected at
optical wavelengths.}

\item{{\bf \object{PKS0945+07}}. This object lies below the
correlation, and is an example of an object with an inverted MFIR
spectrum that apparently lacks a cool dust component
(e.g. \citealp{miley84}, \citealp{vanbemmel01}).}

\item{{\bf \object{PKS1306-09}}. This object was discussed in
\citet{tadhunter02}, and shows marginal polarization at optical
wavelengths, which could be due to scattered light or a non-thermal
optical continuum component. In addition, the extrapolation of the
radio component in the SED (see D08) is consistent with non-thermal
contamination of the thermal MFIR continuum. So far, the optical
spectra are inconclusive regarding the presence of a young stellar
component in this object, but it is difficult to entirely rule out the
presence of such a component.}

\end{itemize}

\clearpage
\begin{deluxetable}{c@{\hspace{2mm}}c@{\hspace{0mm}}l@{\hspace{0mm}}l@{\hspace{0mm}}l@{\hspace{0mm}}l@{\hspace{0mm}}c@{\hspace{0mm}}c@{\hspace{0mm}}}
\tabletypesize{\scriptsize}
\tablecaption{The Supplementary Starburst Sample \label{tbl-3}}
\tablewidth{0pt}
\tablehead{
\colhead{Object}{\hspace{2mm}} &  \colhead{z}{\hspace{0mm}} &\colhead{$L_{24}$(W/Hz)}{\hspace{0mm}} & \colhead{$L_{70}$(W/Hz)}{\hspace{0mm}} & \colhead{$L_{[\rm{OIII}]}$(W)}{\hspace{0mm}}& \colhead{$L_{5GHz}$(W/Hz)}{\hspace{0mm}}& \colhead{SB ref}{\hspace{0mm}} & \colhead{[OIII] ref}{\hspace{0mm}}
}
\startdata
3C48	&	0.367	&$	\phantom{aa}	3.0 \times10^{25	}$&$	\phantom{aa}	1.6\times10^{26	}$&$	\phantom{aa}	3.0 \times10^{36	}$&$	\phantom{aa}	 1.6\times10^{26	}$&	6	&	4	\\
3C213.1	&	0.194	&$	\phantom{aa}	5.1\times10^{24	} a$&$	\phantom{aa}	5.0\times10^{24	} a$&$	\phantom{aa}	5.6\times10^{33	}$&$	\phantom{aa}	3.1\times10^{25	}$&	18	&	11	\\
3C236	&	0.101	&$	\phantom{aa}	2.0 \times10^{23	}$&$	\phantom{aa}	5.7 \times10^{23	}$&$	\phantom{aa}	1.8 \times10^{33	}$&$	\phantom{aa}	1.7\times10^{25	}$&	1,5	&	2	\\
3C285	&	0.079	&$	\phantom{aa}	9.3\times10^{23	} a$&$	\phantom{aa}	1.7 \times10^{24	} a$&$	\phantom{aa}	3.1 \times10^{33	} a$&$	\phantom{aa}	4.1\times10^{24	}$&	1,2	&	2	\\
3C293	&	0.045	&$	\phantom{aa}	9.2 \times10^{22	}$&$	\phantom{aa}	6.7 \times10^{23	}$&$	\phantom{aa}	1.3 \times10^{32	}$&$	\phantom{aa}	 4.0\times10^{24	}$&	7	&	1	\\
3C305	&	0.042	&$	\phantom{aa}	9.4\times10^{22	} a$&$	\phantom{aa}	5.1 \times10^{23	} a$&$	\phantom{aa}	1.4 \times10^{34	} a$&$	\phantom{aa}	1.7\times10^{24	}$&	7	&	3	\\
3C321	&	0.096	&$	\phantom{aa}	3.3 \times10^{24	}$&$	\phantom{aa}	1.0 \times10^{25	}$&$	\phantom{aa}	8.8 \times10^{34	}$&$	\phantom{aa}	1.2 \times10^{25	}$&	1,3	&	2	\\
3C433	&	0.102	&$	\phantom{aa}	2.1 \times10^{24	}$&$	\phantom{aa}	3.5 \times10^{24	}$&$	\phantom{aa}	2.2 \times10^{34	}$&$	\phantom{aa}	4.2 \times10^{25	}$&	1,4	&	3	\\
B2 0648$+$27	&	0.041	&$	\phantom{aa}	1.5 \times10^{24	} a$&$	\phantom{aa}	4.6 \times10^{24	} a$&$	\phantom{aa}	1.2 \times10^{34	} a$&$	\phantom{aa}	 8.6\times10^{22	}$&	12	&	7	\\
B2 0722$+$30	&	0.019	&$	\phantom{aa}	1.6 \times10^{23	} a$&$	\phantom{aa}	1.1 \times10^{24	} a$&$	\phantom{aa}	5.6 \times10^{31	} a$&$	\phantom{aa}	 2.4\times10^{22	}$&	13,14	&	8	\\
Cen A	&	3.4Mpc	&$	\phantom{aa}	8.5\times10^{21	} a$&$	\phantom{aa}	1.2\times10^{22	} a$&$	\phantom{aa}	5.0\times10^{33	}$&$	\phantom{aa}	2.4\times10^{23	}$&	15,16	&	9,10	\\
PKS0131$-$36	&	0.030	&$	\phantom{aa}	1.5 \times10^{23	} a$&$	\phantom{aa}	1.6 \times10^{24	} a$&$	\phantom{aa}	2.5 \times10^{32	} a$&$	\phantom{aa}	 3.5\times10^{24	}$&	1	&	5	\\
PKS0320$-$37	&	0.006	&$	\phantom{aa}	7.8 \times10^{21	}$&$	\phantom{aa}	1.5 \times10^{23	}$&$	\phantom{aa}	3.0 \times10^{31	}$&$	\phantom{aa}	 2.5\times10^{24	}$&	8,9	&	5	\\
PKS0453$-$20	&	0.035	&$	\phantom{aa}	1.2 \times10^{23	} a$&$	\phantom{aa}	8.8 \times10^{23	} a$&$	\phantom{aa}	6.9 \times10^{31	} a$&$	\phantom{aa}	2.1\times10^{24	}$&	11	&	5	\\
PKS1345$+$12	&	0.122	&$	\phantom{aa}	1.2 \times10^{25	} a$&$	\phantom{aa}	3.4 \times10^{25	} a$&$	\phantom{aa}	9.0 \times10^{34	} a$&$	\phantom{aa}	4.6\times10^{25	}$&	1,10	&	6	\\
PKS1549$-$79	&	0.152	&$	\phantom{aa}	1.2\times10^{25	} a$&$	\phantom{aa}	2.7\times10^{25	} a$&$	\phantom{aa}	8.9\times10^{34	} a$&$	\phantom{aa}	 1.1\times10^{26	}$&	17	&	5	\\

\enddata

\tablecomments{ Table \ref{tbl-3} supplementary sample comprising of
all radio-loud objects from the literature with spectroscopic evidence
for recent star formation activity at optical wavelengths. All
infrared fluxes were obtained from NED apart from the object 3C305 for
which Spitzer/MIPS fluxes were extracted from the archive images. 24
and 70\moo\ luminosities for 3C213.1 are derived from IRAS upper
limits. SB references: (1) \citet{holt07}, (2) \citet{aretxaga01}, (3) \citet{tadhunter96}, (4) \citet{wills02}, (5) \citet{odea01}, (6)
\citet{canalizo01}, (7) \citet{tadhunter05}, (8) \citet{goudfrooij01}, (9) \citet{kuntschner00}, (10) \citet{rodriguez07},
(11)\citet{wills04}, (12)\citet{emonts06}, (13) \citet{emonts_thesis}, (14) Emonts et al. in preperation, (15) \citet{peng02},
(16) \citet{peng04}, (17) \citet{holt07}, (18) \citet{wills08}. [OIII] flux references:
(1) \citet{emonts05}, (2) \citet{saunders89}, (3) Robinson PhD thesis,
2001, Universtiy of Sheffield, (4) \citet{chatzichristou99},
(5) \citet{tadhunter93}, (6) \citet{gelderman94}, (7) \citet{emonts06},
(8) Emonts private communication, (9) \citet{simpson98}, (10) \citet{storchi97}, (11) \citet{holt_thesis}.}

\tablenotetext{a}{derived from IRAS measurements at 24\moo\ and 60\moo.}
\end{deluxetable}
\clearpage

\subsection{Correlation slope statistics}
In the calculation of our fitted regression line plotted in Figure
\ref{fig:starburst}, we have chosen not to include those objects identified as
having evidence for a starburst component. Using a bootstrap technique
we have investigated the uncertainty in the slopes of our fit to these
correlations and tested the effect that the removal of the 70\moo\
upper limits and starburst objects would have on the regression lines.

In order to investigate the uncertainties, a sample of N data points ($x$ and
$y$ points from Figure \ref{fig:starburst}) were labelled and then
drawn at random to create a group of N replacements for the original
sample. We then recalculated the slope and repeated this process 1000
times. Provided the data points are independent, the distribution of
the slopes estimated in the bootstrap trials provide an indication of
the uncertainty in the estimated slopes.
   
The data fitted for this bootstrap are identical to those used in the
Spearman rank correlation statistics (\S \ref{sec:rankstat}) i.e., with
redshifts limited to z $>$ 0.06 in order to avoid most of the objects
with upper limits in their [\rm{OIII}] luminosity and
PKS1839$-$48. Here we investigate the effects of including starburst
objects and the upper limits in the sample drawn for the N data
replacements, in order to quantify their effect on the fitted slopes.

The starburst objects are simply included or excluded when calculating
the slope however, exactly as the Spearman rank test (\S
\ref{sec:rankstat}), the 70\moo\ upper limits were included by randomly
choosing a ratio of 70\moo/24\moo\ to derive new 70\moo\ upper
limits. This process was repeated for each of the 1000
bootstrap trials. The results of our analysis are presented in Table
\ref{tbl-4}.

Column 2 of Table \ref{tbl-4} contains the values of the slopes we
have fitted to the data in Figure \ref{fig:starburst} (the bisector of
a linear least squares fit of $x$ on $y$ and $y$ on $x$) as well as
the slope values calculated with starburst objects excluded.  Columns
3 and 4 present the mean value for 1000 slopes of the sample, created
using a bootstrap technique, column 3 includes the 70\moo\ upper
limits and column 4 is for the sample bootstrapped without upper
limits.

Considering, in the first instance, the results for the $L_{24\mu m}$
vs. $L_{\rm{[OIII]}}$ correlation, we find that the addition/removal
of the starburst objects and the objects with upper limits in 70\moo\
has little effect on the resulting mean slope values, where all values
are consistent within the estimated 1$\sigma$ uncertainties.

Secondly, looking at the results in Table \ref{tbl-4} for the
$L_{70\mu m}$ vs. $L_{\rm{[OIII]}}$ slopes, it evident, that although
the upper limits do not significantly affect the mean regression line
slopes and associated uncertainties, the introduction of the starburst
sample does. This further supports our hypothesis that the starburst
objects are responsible for much of the scatter in the
$L_{\rm{[OIII]}}$ vs. $L_{70\mu m}$ correlation. Excluding the optical
starburst objects, the slopes and associated uncertainties of the
$L_{70\mu m}$ vs $L_{\rm{[OIII]}}$ correlation are entirely consistent
with those of the $L_{24\mu m}$ vs $L_{\rm{[OIII]}}$ correlation,
reinforcing the idea that the dust radiating at both mid-IR and far-IR
wavelengths has a common heating mechanism: AGN illumination.

Given that the bootstrap sample in column 3 is identical to that used
to plot the regression line in Figure \ref{fig:starburst}, these
statistics map the uncertainty in our slope fit. For the sample with
upper limits but not including the starburst objects, we calculate
that the uncertainties in our regression line slopes are 0.05 and 0.07
for the 24\moo\ and 70\moo\ correlations respectively, from the
standard deviations of the bootstrap slopes.

\begin{deluxetable}{c@{\hspace{0mm}}c@{\hspace{0mm}}c@{\hspace{0mm}}c}
\tabletypesize{\scriptsize}
\tablecaption{Bootstrap Analysis \label{tbl-4}}
\tablewidth{0pt}
\tablehead{
\colhead{Correlation plot}{\hspace{0mm}} & \colhead{Bisector}{\hspace{0mm}} &  \colhead{with 70\moo\ UL}{\hspace{0mm}} &\colhead{without 70\moo\ UL}{\hspace{0mm}}  
}
\startdata
24\moo\ vs [\rm{OIII}] with starburst	&	0.78	&	0.79 $\pm$ 0.05	&	0.76 $\pm$ 0.08	\\
24\moo\ vs [\rm{OIII}] without starburst	&	0.75	&	0.75 $\pm$ 0.05	&	0.71 $\pm$ 0.07	\\
\cutinhead{}							
70\moo\ vs [\rm{OIII}] with starburst	&	0.83	&	0.88 $\pm$ 0.11	&	0.93 $\pm$ 0.13	\\
70\moo\ vs [\rm{OIII}] without starburst	&	0.71	&	0.72 $\pm$ 0.07	&	0.78 $\pm$ 0.09	\\

\enddata

\tablecomments{Table \ref{tbl-4}: Slopes for the complete sample excluding
objects with z $>$ 0.06 and PKS1839$-$48. The uncertainties on the
values presented in columns 3 and 4 are the standard deviations of the
slopes derived from the 1000 bootstrap cycles. }
\end{deluxetable}

\section{Discussion}
\subsection{Heating mechanism}

On the basis of our observational results, we have concluded that the
MFIR continuum emitted by powerful radio galaxies is predominantly a
consequence of AGN heating of the circum-nuclear dust. However, far-IR
emission from dust heated by starbursts is apparent in the minority of
the objects that fall above the main correlations involving 70\moo.

The link between far-IR bright radio sources and optical evidence for
starburst activity has been noted in previous
studies. \citet{tadhunter02} found that the two objects with the
strongest evidence for optical starburst activity in their sample are
also the most luminous in the far-IR. Subsequently
\citet{wills02,wills04}, while investigating the UV excess in radio
galaxies, also found that those objects with the best evidence for young
stars have significantly larger far-IR luminosities than the rest of
their sample. The Spitzer results for this 2Jy sample strongly
reinforce these earlier results with a much larger sample.

\begin{figure}[h]
\epsscale{1}
\plotone{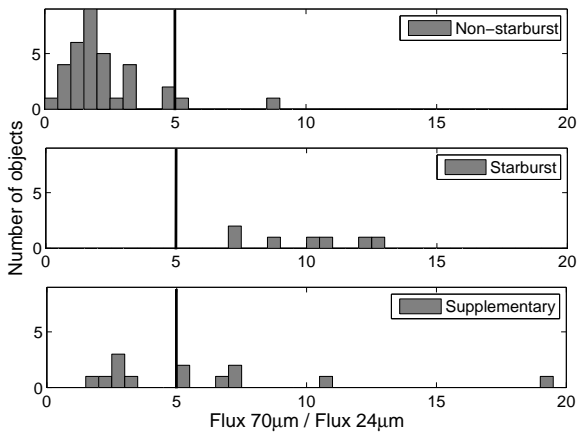}
\caption{Histograms of MFIR colours (70\moo/24\moo\ flux ratio) for
the non-starburst sample, our starburst sample and the supplementary
starburst sample. The black line shows the dividing line between what we define as cool
and warm colours (70\moo/24\moo\ = 5).\label{fig:starburst_hist}}
\end{figure}

Contrary to our results, previous studies such as those by
\citet{rowan95}, \citet{vanbemmel01, vanbemmel03}, put forward
starburst heating as the primary heating mechanism of the dust
producing far-IR emission in powerful AGN, albeit without direct
evidence.  This has encouraged some investigators to search for more
direct observational evidence of a starburst/AGN link. For example,
\citet{schweitzer06} and \citet{netzer07} analyzed PAH starburst
tracers and MFIR emission. Although unable to detect PAH features in
60\% their sample, \citet{schweitzer06} suggest that star formation
heating most likely accounts for at least 30\% of the far-IR emission
of their sample of PG quasars, finding a correlation between PAH
luminosity at 7.7\moo\ and 60\moo\ luminosity similar to that of
starburst dominated ULIRGs. Although \citet{schweitzer06} provide some
of the best empirical evidence for a starburst/AGN connection, albeit
with many upper limits both in the far-IR and PAH emission, it is not
clear that PAH features are exclusively starburst signatures. Indeed,
studies such as that by \citet{siebenmorgen04} have suggested that it
may be possible to produce PAH emission from AGN heating alone if the
dust is situated at sufficiently large radii from the AGN.

We have shown that the far-IR continuum luminosities ($L_{70\mu m}$)
of the 2Jy radio sources are correlated with their [OIII] luminosities
($L_{[OIII]}$) with a similar slope to the correlation between
$L_{24\mu m}$ and $L_{[OIII]}$. The similarity of the two correlations
points towards a common mechanism for producing both the mid and
far-IR emission. If we assume, as is generally accepted, that the
continuum emission at 24\moo\ is heated almost exclusively by the AGN
then, in order for the cool dust emitting in the far-IR to be heated
predominantly by starbursts rather than the AGN, a remarkable degree
of coordination between AGN and starburst activity would be
implied. Although we cannot entirely rule out this possibility, we
regard it as unlikely.

Empirically, starburst galaxies are known to be associated with cool
MFIR colours, as expected in the case of illumination of extended dust
structures by spatially distributed star forming regions. In Figure
\ref{fig:starburst_hist} we present histograms showing the
70\moo/24\moo\ MFIR colours of the non-starburst, starburst and
supplementary samples separately. It is immediately apparent from the
plot that \emph{all} the 2Jy starburst objects have cool colours
($70\mu m/24\mu m > 5$), whereas all but two of the non-starburst
sample have much warmer colours ($70\mu m/24\mu m < 5$). The
non-starburst object in Figure \ref{fig:starburst_hist} with the
coolest colour is PKS 0347+05; as discussed above, it is possible that
this object may contain as yet undetected star formation at optical
wavelengths. Moreover, one of the non-starburst objects whose starburst
status is uncertain -- PKS1306-09 -- also has relatively cool colours
(70\moo/24\moo\ = 4.7).

The data presented in Figure \ref{fig:starburst} can also be used to
consider links between AGN and starburst activity. As discussed in \S
\ref{sec:looking} at low emission line luminosities ($L_{\rm{[OIII]}}
< 10^{33}$W) it is evident that $L_{\rm{[OIII]}}$ is only weakly
correlated with the far-IR luminosity. However, it is notable that all
the objects with large 70\moo\ luminosities ($L_{70\mu m} \geq
10^{25}$W/Hz) --- many of which we have identified with starburst
heating of the cool dust --- also have large [OIII] luminosities
($L_{\rm{[OIII]}} \geq 10^{35}$W). This suggest a weak link between
starburst and AGN activity in the sense that only the most powerful
radio-loud AGN are associated with powerful, ULIRG-like
starbursts. However, the reverse is not true since many of the highest
emission line luminosity objects --- some of which are luminous enough
at MFIR wavelengths to be classified as LIRGs --- do not show evidence
for starburst activity, suggesting that powerful AGN are not always
accompanied by massive star formation activity. Similar trends have
been found in a study of nearby type 2 Seyfert galaxies by
\citet{gonzalez01}, who show that, for a given [OIII] luminosity,
there are two populations of Seyferts: one with optical evidence for
prodigious recent star formation, large far-IR luminosities and cool
MFIR colours; the other with little evidence for enhanced star
formation activity. However, in the case of the Seyferts, a higher
proportion ($>$50\%) of objects show evidence for recent star
formation activity compared with our sample of radio galaxies.

We can also estimate the rate of energetically significant starburst
activity in our sample by considering the main optical and infrared
indicators of starbursts: 7 (15\%) of the objects in the 2Jy sample
show unambiguous spectroscopic evidence for recent star formation
activity at optical wavelengths; 9 (20\%) have cool MFIR colours
($L_{70\mu m}/L_{24\mu m} > 5$); 12 (26\%) of the objects lie more
than 0.3 dex ($\approx$factor$\times$2) about the regression line in
the $L_{70\mu m}$ vs $L_{\rm{[OIII]}}$ correlation in Figure 6; and 13
objects (28\%) show at least one of these indicators. Therefore we
estimate that the proportion of powerful radio-loud AGN showing
evidence for energetically significant recent star formation activity
is in the range 15-28\%.

As first described in \citet{tadhunter07}, this brings us to a key
result: given the lack of a correlation between starburst and AGN
activity, and the fact that only a minority of objects in our sample
show any evidence for recent star formation activity, it is unlikely
that all powerful radio galaxies are triggered at the peaks of major
gas rich mergers. This conclusion is strengthened by the fact that not
all powerful radio galaxies show tidal tails or other evidence for
recent galaxy merger activity (e.g \citealp{heckman86},
\citealp{tadhunter89}, \citealp{mclure99}).

\subsection{Energetic feasibility}

Having found preliminary evidence, based on the correlations between
MFIR and [OIII] luminosities, that the MFIR emitting dust is
predominantly heated by AGN, it is important to assess whether this
heating mechanism is energetically feasible. We start by assuming a
simple model in which the far-IR continuum, mid-IR continuum and
[OIII] emission lines are produced by AGN illumination of structures
with covering factors $C_{fir}$, $C_{mir}$ and $C_{nlr}$
respectively. We make no assumptions about the radial distribution of
dust, although the far-IR emitting dust must be situated at larger
radial distances from the AGN than the mid-IR emitting dust in order
to produce its cooler temperature.

  Based on simple recombination theory, the total $H\beta$ luminosity
($L_{H\beta}$), generated in the narrow line region (NLR) of an AGN,
is related to the ionising luminosity ($L_{ion}$) by

\begin{displaymath}
{L_{ion}} = L_{H\beta}\frac{\langle h\nu \rangle_{ion}}{h\nu_{H\beta}} \frac {\alpha^{B}_{eff}}{\alpha_{H\beta}^{B}} C_{nlr}^{-1}
\end{displaymath}

\noindent where $\langle h\nu \rangle_{ion}$ is the mean ionising
photon energy, $h\nu_{H\beta}$ is the energy of an $H\beta$ photon,
$\alpha^{B}_{eff}$ is the total case B recombination coefficient and
$\alpha_{H\beta}^{B}$ is the effective $H\beta$ recombination coefficient.  From
the results of \citet{elvis94} the bolometric luminosity $L_{bol}$ is
related to the ionising luminosity by $ L_{bol} \approx 3.1 L_{ion}$,
therefore

\begin{displaymath}
{L_{H\beta}} = 0.32 L_{bol}\frac{h\nu_{H\beta}}{\langle h\nu \rangle_{ion}} \frac {\alpha_{H\beta}^{B}}{\alpha^{B}_{eff}} C_{nlr}.
\end{displaymath}

At high luminosities and ionization parameters, $L_{[\rm{OIII}]}
\approx 12 L_{H\beta}$. In addition, fits to the line ratios of nearby
radio galaxies are consistent with $\langle h\nu \rangle_{ion}
$=$6.2\times 10^{-11}J$, corresponding to an ionising continuum shape
with $\beta = 1.5$ ($F_{\nu} \propto \nu^{-\beta}$)
\citep{robinson87}. Also, given that $\alpha^B_{H\beta}/\alpha_{eff} = 0.09$
\citep{osterbrock}, we find:

\begin{equation}
L_{[\rm{OIII}]} = 2.2 \times 10^{-2} L_{bol} C_{nlr}.
\end{equation}

\noindent This result can be related to the MFIR luminosities by assuming that the
structures producing the MFIR absorb the short wavelength radiation
and re-radiate it in the MFIR, then $L^{bol}_{mir} = C_{mir}L_{bol}$,
and similarly $L^{bol}_{fir} = C_{fir}L_{bol}$. Defining the
wavelength range of the mid-IR as 2 -- 30\moo, and that of the far-IR
as 30 -- 100\moo, the mid-IR and far-IR bolometric luminosities are
defined as:

\begin{displaymath}
L_{mir}^{bol} = \int_{1\times10^{13}Hz}^{1.5\times10^{14}Hz} L_{\nu} d\nu
\end{displaymath}

and

\begin{displaymath}
L_{fir}^{bol} = \int_{3\times10^{12}Hz}^{1\times10^{13}Hz} L_{\nu} d\nu.
\end{displaymath}

Representing $L_{\nu}$ as a power law $L_{\nu}=K\nu^{\gamma}$, we can then write 
\begin{displaymath}
L_{mir}^{bol} = \int_{1\times10^{13}Hz}^{1.5\times10^{14}Hz} K\nu^{\gamma} d\nu 
\end{displaymath}

\noindent and

\begin{displaymath}
K_{mir} = \frac{(\gamma+1)L_{bol}C_{mir}}{[\nu^{\gamma+1}]_{1\times10^{13}Hz}^{1.5\times10^{14}Hz}},  
\end{displaymath}

\noindent and we can write a similar expression for the
$K_{fir}$. Given our assumption of a simple power-law shape for the MFIR SED, the
MFIR spectral index is directly calculated from the MFIR colour:

\begin{displaymath}
\gamma=\frac{log (L_{70}/ L_{24})}{log(24/70)}.
\end{displaymath}

\noindent For our complete sample we find a median MFIR colour of
$L_{70\mu m}/L_{24\mu m}=2.1$, leading to $\gamma_{median}\approx
0.7$. Therefore the MFIR monochromatic luminosities can be related to
the AGN bolometric luminosities by:

\begin{displaymath}
L_{mir}= 3\times10^{-5} L_{bol} C_{mir} \nu^{-0.7}
\end{displaymath}
and
\begin{displaymath}
L_{fir}= 1.25\times10^{-4} L_{bol} C_{fir} \nu^{-0.7}.
\end{displaymath}

By substituting equation (1) into these expressions we find:

\begin{equation}
L_{(24\mu m)}=1\times10^{-12} L_{[\rm{OIII}]} \frac{C_{mir}}{C_{nlr}}
\end{equation}
and
\begin{equation}
L_{(70\mu m)}=9\times10^{-12} L_{[\rm{OIII}]} \frac{C_{fir}}{C_{nlr}}.
\end{equation}

\begin{figure*}[t]
\epsscale{2}
\plotone{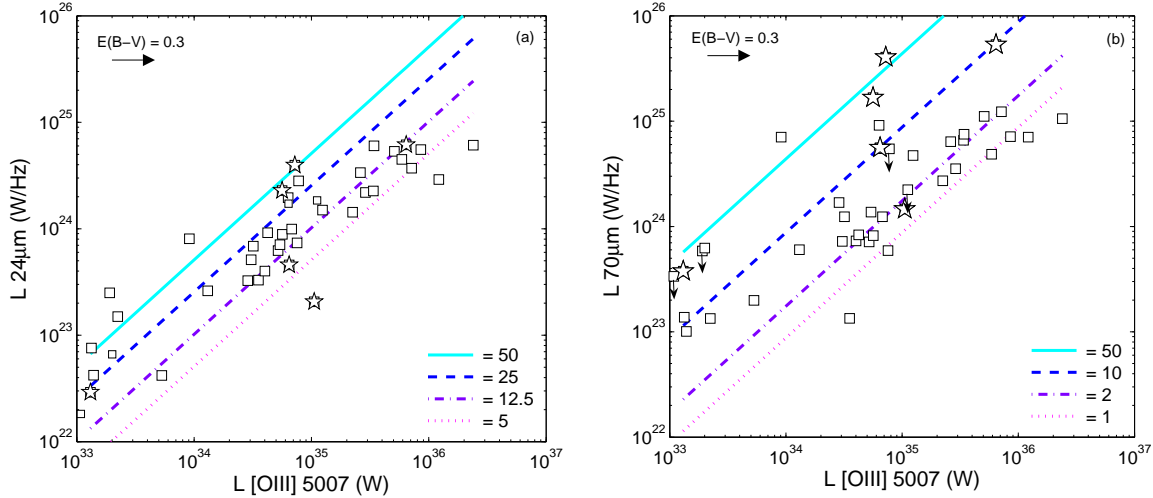}
\caption{Plots of the $L_{[\rm{OIII}]}$ vs.$L_{24\mu}$ (left) and
$L_{[\rm{OIII}]}$ vs.$L_{70\mu}$ (right) for the complete 2Jy sample
with lines showing the range of possible ratios of covering factors
required to explain the trends in terms of the AGN illumination
model. $C_{mir}/C_{nlr}$ in plot (a) and $C_{fir}/C_{nlr}$ in plot
(b). Note our calculation of the covering factors does not seek to
explain the non-linear slope (discussed further in \S
\ref{sec:slopes}). Also the ranges of covering factor ratios plotted
are different in plots (a) and (b). The arrow in the upper left hand
corners of the plots show the effect that correcting for an intrinsic
extinction of E(B-V) = 0.3 would have on the [OIII] emission line
luminosities.
\label{fig:covering}}
\end{figure*}

\noindent It is likely that the major uncertainty in $C_{mir}/C_{nlr}$
is associated with the assumed $ L_{bol}/L_{ion}$ ratio that may be
uncertain by a factor of $\approx 2$ \citep{elvis94}.

 In Figure \ref{fig:covering} we have plotted the $L_{70\mu m}$ vs
$L_{\rm{[OIII]}}$ and $L_{24\mu m}$ vs $L_{\rm{[OIII]}}$ correlations
with predictions from our calculations for a range of
$C_{mir}/C_{nlr}$ and $C_{fir}/C_{nlr}$ covering factor ratios. This
allows us to investigate whether the MFIR emission can be explained
solely in terms of AGN illumination of the dust. Note that our calculations
do not take into account the fact that the $L_{MFIR}$ vs $
L_{[\rm{OIII}]}$ relationships are non-linear (discussed further in \S
\ref{sec:slopes}); the calculated ratio slopes are most relevant to
the high luminosity ends of the correlations. Also, in the calculation
so far we have assumed that the NLR, the mid-IR and far-IR emitting
regions represent discrete structures at different locations in each
galaxy, whereas, in fact, it is more likely that there exists a
continuous distribution from one region to the next, or an overlap
between the regions.

Firstly, Figure \ref{fig:covering}(a) shows that $C_{mir}/C_{nlr}
\approx$ 5-25 would account for the mid-IR luminosities of
the majority of the objects in our sample. However, a value of
$C_{mir}/C_{nlr} \approx 12.5$ provides the best description of the
high luminosity end of the main correlation (from Figure
\ref{fig:starburst}). Therefore NLR clouds
cannot be responsible for the bulk of the mid-IR emission of the radio
galaxies in our sample, a result consistent with detailed models for the mid-IR
emission from NLR clouds \citep{groves06}. We further note that much
of the scatter in the $L_{24\mu m}$ vs $L_{\rm{[OIII]}}$ correlation
can be explained in terms of variations in $C_{mir}/C_{nlr}$.

The absolute covering factor of the dusty torus structure emitting the
mid-IR continuum can be estimated by assuming that the NLR has a
covering factor in the range $C_{nlr} \approx $ 0.02-0.08
(\citealp{netzer93}, \citealp{maiolino01}). For $C_{mir}/C_{nlr}
\approx 12.5$ this gives $0.25<C_{mir}<1$. A large $C_{mir}$ is
entirely feasible in the context of obscuration by the putative dusty
torus required by the orientation-based unified schemes
\citep{barthel89}, since a torus with a typical opening half angle in the
range $45^{\circ}-60^{\circ}$ would have a covering factor of
0.5-0.7. This calculation also indicates that the covering factor of
the NLR must be small ($C_{nlr}<0.08$) if we require
$C_{mir}<$1. Therefore, it is evident that the 24\moo\ continuum
emission from the warm dust component can be feasibly powered by AGN
heating alone for most of the objects; the relative covering factor of
the mid-IR emitting structures is consistent with that expected for a
circum-nuclear torus, which is not itself expected to radiate
significant forbidden line radiation due to high densities and
temperatures.

Secondly, it is clear from Figure \ref{fig:covering}(b) that the ratio
of covering factors required for the emitting dust structures is much
less in the far-IR than at mid-IR wavelengths: typically
$C_{fir}/C_{nlr} \approx 1-2 $ for the main non-starburst sample,
corresponding to $C_{fir} = 0.02-0.16$. Such covering factors are
feasible if cool dust is associated with the outer parts of the torus,
a kpc-scale dust lane such as those found in many nearby radio
galaxies \citep{dekoff00}, or indeed the NLR itself. In the latter
context we note that, if we correct the emission line luminosities for
the reddening typical of the NLR of radio galaxies
(0.3$<$E(B-V)$<$1.0: \citealp{tadhunter94}, \citealp{robinson00},
\citealp{robinson_thesis})\footnote{These reddening estimates are based on
Balmer line ratio measurements obtained after careful subtraction of
the underlying stellar continuum.} this will shift many of the points
in Figure \ref{fig:covering}(b) close to the $C_{fir}/C_{nlr}=1$
line. Moreover, the NLR in many nearby radio galaxies and PG quasars
are extended on sufficiently large spatial scales ($\sim$1kpc or
larger: \citealp{bennert02}, \citealp{privon08}) that the dust
associated with them would be heated by the AGN to the relatively cool
temperatures required to produce the far-IR radiation.  Therefore,
rather than being separate entities, it is entirely plausible that the
NLR clouds \emph{are} the dusty structures emitting the far-IR
continuum. 

We have again marked the starburst objects from our sample in Figure
\ref{fig:covering}. Many of these starburst objects require
$C_{fir}/C_{nlr} \approx 10-50$. If we attempt to explain the far-IR
emission in such objects in terms of AGN illumination alone, this
raises the question of how the AGN could illuminate the far-IR
emitting regions without producing substantial emission line radiation
from the gas associated with the dusty structures (i.e. leading to
$C_{fir}/C_{nlr} = 1 $). One explanation is that the $C_{nlr}$ is
underestimated, and $C_{fir}/C_{nlr}$ overestimated, due to dust
extinction of the NLR. However, this would require more substantial
reddening (E(B-V) $>$ 1) than is supported by most observations of the
NLR. Therefore the objects that lie above the $C_{fir}/C_{nlr}$= 1 line
are likely to have their far-IR flux boosted by starburst heating or
non-thermal emission. Incidentally, this includes \object{PKS0347+05}
and \object{PKS1306-09}, which we discussed in \S \ref{sec:looking} as
possible starburst candidates.

\subsection{Correlation slopes}
\label{sec:slopes}
The fitted regression lines from Table \ref{tbl-4} (0.75 $\pm0.05$ at
24\moo\ and 0.72 $\pm0.07$ at 70\moo) are plotted on Figure
\ref{fig:starburst}. Taking into account the bootstrap uncertainties,
the slopes of the correlations are significantly non-linear. It is
notable that \citet{maiolino07} find a similar slope (0.82$\pm0.02$)
between continuum luminosity at 5100A and infrared luminosity at
6.7\moo\ for a sample of 25 high luminosity QSOs. We now consider
possible explanations for the non-linear slopes.

First we consider explanations related to the emission line physics of
the NLR clouds. The most basic AGN photoionization models assume that
the emission-line regions are photoionised by the AGN and that the
properties of the emission line region, such as density and covering
factor, do not change significantly with AGN power. Considering such a
model for a simple optically thick slab, \citet{tadhunter98} found
that at high luminosities, the relation between $L_{\rm{[OIII]}}$ and
ionising luminosity can be represented as a power-law $L_{ion} \propto
(L_{\rm{[OIII]}})^{\alpha}$, where $\alpha = 0.8\pm0.1$. Assuming that
the properties of the torus do not change with luminosity (but see
below) and that the MFIR continuum luminosity is directly proportional
to $L_{bol}$ and $L_{ion}$, we therefore expect $L_{MFIR}\propto
L_{\rm{[OIII]}}^{\alpha} \propto L_{\rm{[OIII]}}^{0.8\pm0.1} $. This
result is consistent with the slopes of our $L_{MFIR}$ vs
$L_{\rm{[OIII]}}$ correlations as well as that found by
\citet{maiolino07}. However, based on several detailed studies of the
emission line spectra of radio galaxies (e.g. \citealp{robinson00},
\citealp{taylor03}), it is likely that the single slab
photoionization models are over-simplistic, and that multi-component
models for the NLR are more appropriate. In this case the relationship
between $L_{\rm{[OIII]}}$ and $L_{ion}$ is less clear. Moreover,
investigations of the relationship between $L_{\rm{[OIII]}}$ and the
continuum luminosities in large samples of AGN provide direct
observational evidence for steeper slopes in the relationship between
$L_{bol}$ and $L_{\rm{[OIII]}}$, albeit with larger
scatter ($\alpha \geq 1$ \citealp{netzer04, netzer06}).

Alternatively, we can also consider whether the non-linear slopes are
consistent with the basic receding torus model that has been suggested
to explain the variation in the BLRG+Q/NLRG fraction with luminosity
in the context of the unified schemes (\citealp{lawrence91}, \citealp{hill96},
\citealp{simpson98, simpson05}). This solution is attractive because it is widely
accepted that, if the mid-IR emitting dust lies close to the AGN, the
inner radius of the torus at which the dust is sublimated due to
illumination by the AGN must depend on AGN power. From simple thermal
equilibrium arguments we can relate the distance of the illuminating
AGN from the inner face of the torus (r) to the bolometric luminosity
($L_{bol}$) and the sublimation temperature of the dust ($T_{sub}$) as
follows:

\begin{equation}
r \propto \left(\frac{L_{bol}}{T_{sub}^4}\right)^{1/2}.
\end{equation}

The area of the inner face of the obscuring torus can be written as $
4\pi rh$ where $h$ is the thickness of the torus from its mid-plane to its
top surface. The covering fraction of the torus is then:

\begin{equation}
C_{tor} = \frac{4\pi rh}{4\pi r^2} = \frac{h}{r},
\end{equation}

assuming all the radiation absorbed by the face of the torus is
re-emitted by dust, and $h$ is fixed, then the relationship between
emission in the mid-IR at 24\moo\ emitted by dust close to the AGN and
the bolometric luminosity can be written as:

\begin{displaymath}
P_{24\mu m} \propto \frac{h}{r}\, L_{bol} \propto  L_{bol}^{1/2}.
\end{displaymath}

\noindent having substituted equation (5) into equation (4). Finally, assuming that $L_{bol} \propto
\left(L_{[\rm{OIII}]}\right)^{\alpha}$ we obtain:

\begin{displaymath}
P_{24\mu m} \propto \left(L_{[\rm{OIII}]}\right)^{\alpha/2}.
\end{displaymath}

In the context of the receding torus model we now consider three cases with different
relationships between $L_{\rm{[OIII]}}$ and $L_{bol}$:

\begin{enumerate}

\item {\emph{The relationship between $L_{\rm{[OIII]}}$ and $L_{bol}$ is
linear with $\alpha$=1, and therefore the non-linear slopes of
$L_{\rm{[OIII]}}$ vs $L_{MFIR}$ correlations are entirely due to the
receding torus.} In this case $P_{24\mu m} \propto
\left(L_{[\rm{OIII}]}\right)^{0.5}$ which is too shallow to account
for the measured correlation slopes, even when considering the
uncertainties on the measured slope we have estimated from the
bootstrap analysis ($\pm$ 0.05). \citet{netzer04} found evidence for
such a linear relationship between $L_{\rm{[OIII]}}$ and $L_{AGN}$,
from the lack of variation of the equivalent width of [OIII] as a function of
continuum luminosity. However, there are many upper limits for the
high luminosity objects in their sample. }

\item{\emph{The correlation between $L_{\rm{[OIII]}}$ and $L_{bol}$ is not
linear and has a slope $\alpha<1$, as predicted by simple AGN
photoionization models \citep{tadhunter98}}. If $\alpha$=0.8, as
discussed above, then $P_{24\mu m} \propto
\left(L_{[\rm{OIII}]}\right)^{0.4}$. This value is also significantly
shallower than measured from our data.}

\item{\emph{The correlation between $L_{\rm{[OIII]}}$ and $L_{AGN}$ is
non-linear and $\alpha > 1$}. In this context it is notable that
\citet{netzer06} find evidence for $\alpha \approx 1.7\pm0.07$, based
on an analysis of X-ray and emission line properties of luminous
AGN. This then gives $P_{24\mu m} \propto
\left(L_{[\rm{OIII}]}\right)^{0.85\pm0.04} $, which is consistent with the
measured slopes within the uncertainties. }

\end{enumerate}

Option 1, considering only a receding torus model, cannot
satisfactorily explain our results. \citet{simpson05} also shows that,
assuming isotropic [OIII] emission, the fraction of BLRG/Q increases
with AGN luminosity, but with a shallower slope than predicted by a
pure receding torus model. He proposes an increase in torus height
with luminosity to account for the discrepancy. Although this may not
be the only explanation for the observed trends, the \citet{simpson05}
result shows that there remain questions about the application of the
most basic receding torus model.

In this study we have shown that the slope of the $L_{70\mu m}$ vs
$L_{\rm{[OIII]}}$ correlation is similar to that of the $L_{24\mu m}$
vs $L_{\rm{[OIII]}}$ correlation. If the far-IR radiation is emitted
by extended cool dust structures it is clear that these must lie far
from the AGN for the typical luminosities of our sources
($>0.1$kpc). In this case it is unlikely that the dust structures will
recede in the same way as predicted for the torus. Therefore, while a
receding torus might provide a plausible explanation for the
non-linearity of the $L_{24\mu m}$ vs $L_{\rm{[OIII]}}$ correlation
(for $\alpha>1$) such an explanation is unlikely to apply to the
$L_{70\mu m}$ vs $L_{\rm{[OIII]}}$ correlation.
 
\subsection{Unification}
\label{sec:unify_disc}

Given the completeness of our sample, the similarities between the
MFIR properties of the NLRG and BLRG/Q classes provide excellent
support for the orientation-based unified schemes for radio-loud
AGN. In addition, our results provide an important constraint for
understanding the structure and optical depth of the obscuring
torus. The question then remains: how do we reconcile our results with
previous studies that did find evidence for significant obscuration at
MFIR wavelengths? Apart from sample selection and incompleteness (see
\S \ref{sec:intro}), there are several factors that may have biased
previous studies and led to the finding of a difference between the
two classes of radio-loud AGN:

\begin{itemize}

\item{{\bf Classification of BLRG objects}: The deep spectra available
for all our sample objects allow us to make secure classifications of
objects into NLRG, BLRG/Q and WLRG categories. Past studies may have
only included those objects with the most luminous broad lines;
objects with weaker broad lines may have been missed. Therefore,
excluding the weak BLRG from the broad line samples would have biased
BLRG/Q towards higher luminosities compared with the NLRG.}

\item{{\bf Weak line radio galaxies}: From Figure \ref{fig:opradall}
(a) \& (b), it is clear that the WLRG lie at the low luminosity end of
both the MFIR bands. Many previous studies have not made the
distinction between NLRG and WLRG objects. In not recognising these
radio-loud galaxies as a separate population, it is clear that the
inclusion of WLRG with NLRG would lower the mean MFIR luminosity of
samples of NLRG/WLRG, making the BLRG/Q appear relatively stronger
emitters at MFIR wavelengths (see \citealp{laing94} for a a similar
discussion relating to the [OIII] luminosities).}

\item{{\bf Beamed non-thermal MFIR components}: When considering the
MFIR emission from radio-loud galaxies, it is important to quantify
the degree of non-thermal contamination of the thermal infrared
emission from dust. Past investigations that did not account for such
contamination (e.g. including flat spectrum objects) may have suffered
a bias in their results. Contamination from a beamed non-thermal
component is more likely in BLRG/Q oriented close to the line of
sight, and therefore could have boosted the MFIR luminosities of some
of the BLRG/Q objects in previous studies. As mentioned in Section
\ref{sec:OpClass} we find no more than 24\% of our 2Jy sample have a
possibility of contamination from non-thermal beamed components (see
\citet{dicken08}), consistent with the results presented in
\citet{cleary07}.}
\end{itemize}

Although the results of our study support the unified schemes, further
data are required at shorter wavelengths ($\lambda < 24 \mu m$) to
provide necessary evidence for greater optical depth in NLRG compared
with BLRG/Q objects. Additionally, our results have been based on the
assumption that the [OIII] emission line luminosity is isotropic, and
it is important to consider whether this assumption is valid.

\begin{figure}[t]
\epsscale{1}
\plotone{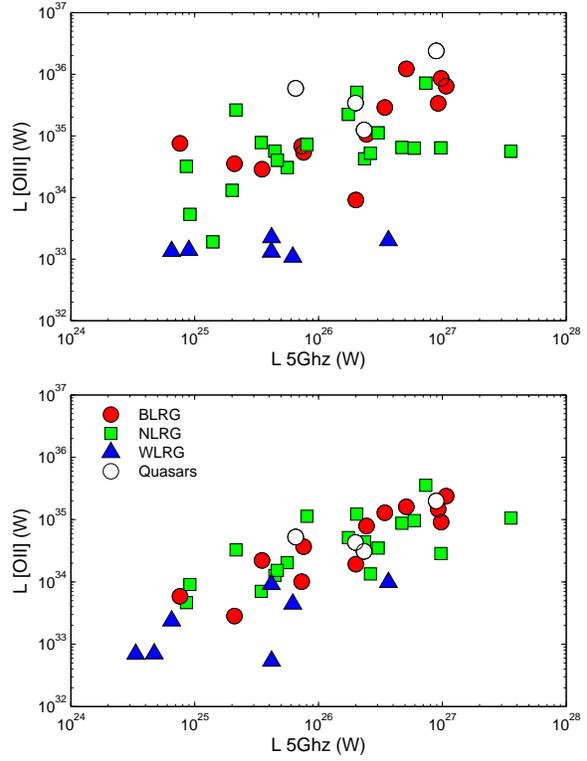}
\caption{Plots of [OIII]$\lambda$5007 and [OII]$\lambda$3727 vs 5 \rm{GHz} radio total radio
luminosity. Not including 4 upper limits in [OIII] and 5 upper
limits in [OII] \label{fig:opOIIIRad}}
\end{figure}

Previous studies have presented evidence that quasars
have higher [OIII]$\lambda$5007 luminosities than radio galaxies
(\citealp{jackson90}, \citealp{haas05}), similar to the results in
opposition to BLRG/Q and NLRG unification in the MFIR
(\citealp{heckman94}, \citealp{hes95}). In fact, returning to Figure
\ref{fig:opradall}, it is notable that there is a weak tendency for
the BLRG/Q sources to have higher [OIII]$\lambda$5007 luminosities
than the NLRG sources for a given MFIR luminosity. On the other hand, \citet{hes93, hes96}
demonstrate that the [OII]$\lambda3727$ line, with its lower
ionization potential and critical density, shows no evidence for a
difference in luminosity between BLRG/Q and NLRG.

\begin{figure}[t]
\epsscale{1}
\plotone{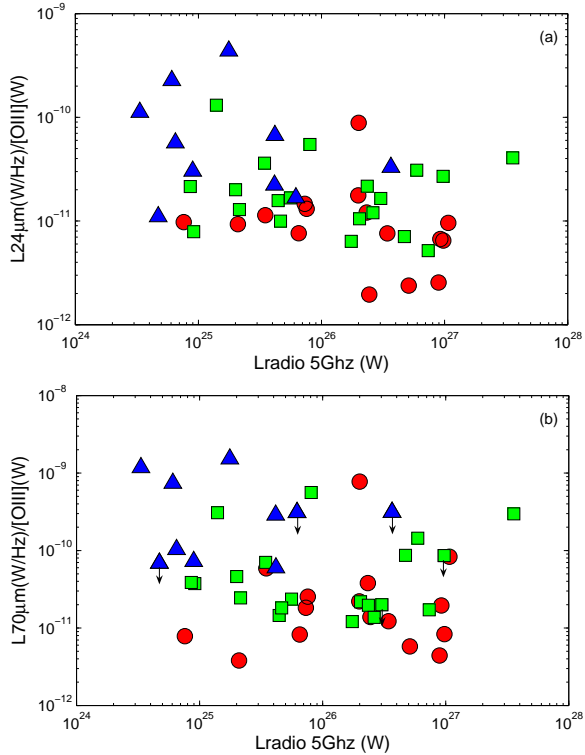}
\caption{Plots of 24\moo\ (top) and 70\moo\ (bottom) luminosities normalized by [OIII]$\lambda$5007 luminosity vs 5 \rm{GHz} radio total radio
luminosity. \label{fig:Rdr}}
\end{figure}

In Figure \ref{fig:opOIIIRad} we present $L_{\rm{[OIII]}\lambda5007}$
and $L_{\rm{[OII]}\lambda3727}$ plotted against total radio luminosity
($L{5GHz}$). A visual inspection of this figure reveals that the
distributions of the NLRG and BLRG/Q are similar across several orders
of magnitudes. Despite the BLRG/Q having some of the highest
luminosities in [\rm{OIII}] emission lines, there are no significant
differences between the two populations in this plot. This is
consistent with the findings of \citet{laing94}, \citet{mulchaey94},
\citet{tadhunter98} and \citet{jackson97}.

In Figure \ref{fig:Rdr} we again present $L_{24\mu}$ and $L_{70\mu}$
plotted against $L_{5\rm{GHz}}$, however in this Figure we have
normalized the MFIR luminosities by $L_{\rm{[OIII]}\lambda5007}$. If
we make the assumption that both the [OIII] and MFIR emission are
isotropic then normalising the MFIR luminosity by
$L_{\rm{[OIII]}\lambda5007}$ should reveal no difference in the
distribution of the different optical classifications in the
plot. However, from a visual inspection of Figure \ref{fig:Rdr}, it is
clear that, although there is a large scatter, the BLRG/Q objects
clearly tend towards lower MFIR/[OIII] ratios than NLRG; the three
lowest $L_{24\mu}$/$L_{\rm{[OIII]}\lambda5007}$ ratios, and six lowest
$L_{70\mu}$/$L_{\rm{[OIII]}\lambda5007}$ ratios are associated with
BLRG/Q. The BLRG/Q clearly tend to lower normalised MFIR luminosities
than the NLRG. There are three explanations for such a result.

\begin{itemize}

\item{ The [OIII] emission is isotropic, but the MFIR emission is
weaker for a given [OIII] luminosity in the BLRG/Q objects than it is
in the NLRG objects. This explanation is difficult to square with the
orientation-based unified schemes. In particular, the optical [OIII]
emission is more likely to suffer dust obscuration effects due to the
circumnuclear torus than the MFIR emission. Moreover, assuming that
their jets are truly pointing closer to the line of sight, the BLRG/Q
are less likely to be affected by dust obscuration than the NLRG.
Note, however, that it is possible that some individual BLRG/Q may
have intrinsically under-luminous MFIR emission (e.g. PKS1932-46: see
\citealp{inskip07}).}

\item{ The [OIII] emission is affected by torus dust obscuration and
therefore appears stronger in the BLRG/Q than it is in the NLRG for a
given isotropic MFIR luminosity. This explanation is certainly
consistent with the orientation-based unified schemes, and the
conclusions of various other studies (e.g. \citealp{jackson97},
\citealp{haas05}). It is also consistent with the observational
evidence for narrow line variability on a timescale of years in the
BLRG 3C390.3 (\citealp{clavel87}, \citealp{zheng95}), which suggests
that a proportion of the narrow [OIII] emission may be emitted on a
sufficiently small scale to be affected by torus obscuration. On the
basis of Figure 10, the degree of anisotropy in [OIII] is a factor
$\sim$2 -- 3 for the majority of the objects, only reaching a factor
of 10 in a few extreme cases.}

\item{The [OIII] emission is isotropic but the
$L_{MFIR}/L_{\rm{[OIII]}}$ ratio decreases with luminosity because the
covering factor of the torus decreases as the torus recedes. The
receding torus model predicts that the BLRG/Q objects will have higher
[OIII] luminosities on average than NLRG, even in the absence of dust
obscuration, provided that there exists a range of intrinsic AGN
luminosities for a given isotropic radio power. This is because the
objects hosting the most intrinsically powerful AGN have a greater
chance of being detected as BLRG/Q. As the inner face of the torus
recedes with the luminosity the covering factor of the torus will
decrease. This will lead to the following dependence of the
$L_{MFIR}/L_{\rm{[OIII]}}$ ratio on $L_{bol}$ at fixed radio power:
$L_{MFIR}/L_{\rm{[OIII]}} \propto L^{1/2}_{bol}$ (See Section
\ref{sec:slopes}) assuming a linear relationship between
$L_{\rm{[OIII]}}$ and $L_{bol}$. Therefore in order to explain the
BLRG/Q $L_{MFIR}/L_{\rm{[OIII]}}$ ratios that are a factor $\approx3\times$
lower than the average NLRG, we would require the BLRG/Q to be
intrinsically more luminous than the NLRG at a given radio power by a
factor of $\approx 10$.  However as discussed in Section
\ref{sec:slopes} it is unlikely that this explanation can apply to the
$L_{70\mu m}/L_{\rm{[OIII]}}$ ratio because the structures emitting
the far-IR continuum are unlikely to recede as envisaged by the
receding torus model.}
\end{itemize}

\section{Summary}

We have presented an analysis of deep Spitzer (MIPS) observations for
a complete subset of the 2Jy sample. The main results are as follows:

\begin{enumerate}
\item{{\bf Heating mechanism}: We find tight correlations between the
MFIR luminosities and the [\rm{OIII}]$\lambda5007$ luminosities, with
similar slopes for both 24 and 70\moo\ correlations (0.75$\pm0.05$ at
24\moo\ and 0.72$\pm0.07$ at 70\moo). Given that [\rm{OIII}] is an
indicator of intrinsic AGN strength, we conclude that direct AGN
illumination is the primary heating source for the dust producing both
the mid-IR and far-IR continuum. These correlations are better than
those between MFIR and radio luminosities, and between radio and
[OIII] luminosities. }

\item{{\bf Energetics and dust geometry}: Using simple
arguments we have demonstrated that heating of MFIR-emitting dust
structures by AGN illumination is energetically feasible. We identify
the dust structure producing the mid-IR continuum with the
circum-nuclear torus, and the dust structure producing far-IR
continuum with the NLR clouds. }

\item{{\bf Slopes of the correlations}: Our finding of a non-linear slope in
the $L_{MFIR}$ vs. $L_{[OIII]}$ correlation is quantitatively in support of the
simple receding torus model; however, this result can also be explained
in terms of a non-linear correlation between the strength of [OIII]
optical emission line luminosity and the AGN ionising
luminosity.}

\item{{\bf Unified schemes}: Contrary to some previous studies we find
no strong evidence for increased obscuration of MFIR emission of NLRG
compared to BLRG/Q; the MFIR luminosities of NLRG and BLRG/Q also cover
a similar range. These results fully support the orientation-based
unified schemes, and we hypothesise that previous studies may have been affected
by incomplete, heterogeneous, meager and/or low S/N data.}

\item{{\bf Weak Line radio galaxies}: The WLRG in
our sample are found to have weak MFIR continuum emission as well as
weak [OIII] emission line emission. This implies that the AGN in WLRGs
are intrinsically low luminosity objects, and
that their weak optical emission lines are not simply a consequence of
enhanced dust obscuration by circum-nuclear dust.}

\item{{\bf Starburst heating and triggering}: We have analysed the
increased scatter in the distribution of $L_{70\mu m}$
vs. $L_{[\rm{OIII}]}$ compared to the $L_{24\mu m}$
vs. $L_{[\rm{OIII}]}$ correlation. We find that the increased scatter
is due, in part, to enhanced heating by starbursts in the objects in
our sample that show evidence for recent star formation activity at
optical wavelengths. The relatively low incidence of energetically
significant starburst activity in our sample (15-28\%) has
implications for our understanding of the triggering and evolution of
radio-loud AGN. In particular, only a minority are likely to be
triggered close to the peak of major gas rich mergers with associated
starburst activity. Clearly, a variety of triggering mechanisms must
be present.}

\item{{\bf Anisotropy of [OIII] emission}: Finding lower ratios of
MFIR/[OIII] of BLRG/Q compared to those of NLRG, we have concluded
that the most likely explanation for this result is that there is mild
anisotropy (typically factor $\approx$2-3) in the [OIII] emission due
to obscuration by the circumnuclear dust in the NLRG that is not seen
in the BLRG/Q.}

\end{enumerate}

\acknowledgments 
This work is based [in part] on observations made
with the Spitzer Space Telescope, which is operated by the Jet
Propulsion Laboratory, California Institute of Technology under a
contract with NASA. This research has made use of the NASA/IPAC
Extragalactic Database (NED) which is operated by the Jet Propulsion
Laboratory, California Institute of Technology, under contract with
the National Aeronautics and Space Administration. Based on
observations made with ESO Telescopes at the Paranal Observatory.

DD, JH and KJI acknowledge support from the STFC. KJI acknowledges
support from the German DFG under grant JA1114/3-1 within the Emmy
Noether programme.

{\it Facilities:} \facility{Spitzer (MIPS)}, \facility{ATCA}, \facility{VLA}.

\bibliographystyle{apj.bst} 
\bibliography{bib_list}

\end{document}